\documentclass[aps,a4paper, preprint, superscriptaddress,preprintnumbers,floatfix,nofootinbib,amsmath,amssymb]{revtex4}
\usepackage{epsfig}  
\usepackage{graphicx}
\usepackage{hyperref}
\usepackage{color}
\usepackage{float}
\usepackage{amsfonts}
\usepackage{amsmath}
\usepackage{slashed}
\usepackage{soul}
\large

\begin{document}

\title{Investigating the effect of $U_1$ vector leptoquark on $b\to u \tau\bar{\nu}$ mediated $B$ decays}

\author{Suman Kumbhakar}
\email{suman@phy.iitb.ac.in}
\affiliation{Department of Physics, Indian Institute of Technology Bombay, Mumbai-400076, India}

\author{Rukmani Mohanta}
\email{rmsp@uohyd.ac.in}
\affiliation{School of Physics, University of Hyderabad, Hyderabad-500046, India}

\begin{abstract}
The recent measurements of lepton flavor university (LFU) violating observables in semileptonic $b \to c \ell \bar{\nu} $ and $b\to s\ell^+ \ell^-$ transitions by various experiments  exhibit $(2-3)\sigma$ deviations from their corresponding Standard Model (SM) predictions. These tantalizing signals  hint towards the possible role of new physics (NP) in $b \to c \tau \bar{\nu}$ and $b\to s\mu^+\mu^-$ decay channels. This in turn indicates that the same class of NP  as appeared  in $b \to c \tau \bar{\nu}$, might also show up in other tree level  processes involving   $b \to u \tau \bar{\nu}$ transition. Since these charged current transitions are doubly Cabibbo suppressed, the NP contributions could be significant enough leading to sizeable effects in some of the observables.   In this paper, we study the implications of the vector leptoquark $U_1(3,1,2/3)$, which is one of the few scenarios that can simultaneously explain the LFU violation signals both in the charged-current as well as neutral-current sectors,  on the semileptonic decays $B \to (\pi, \rho, \omega)\tau \bar{\nu}$ and $B_s \to (K, K^*) \tau \bar{\nu}$. In particular, we pay our attention to the branching fraction, lepton flavor non-universality (LNU)  observable, forward-backward asymmetry and the polarization asymmetry parameters of these modes. We find substantial deviations in the branching fractions as well as LNU observables of these decay modes due to the $U_1$ contributions, which can be probed by the currently running experiments LHCb and Belle-II. 
\end{abstract}
 
\maketitle 

 %%%%%%%%%%%%%%%%%%%%%%%%%%%%%%%%%%%%
\section{Introduction} 
%%%%%%%%%%%%%%%%%%%%%%%%%%%%%%%%%%%%%%%%
The SM  is a highly successful and well established theory beyond doubt and can explain almost all the observed data from the colliders.  
Though the LHC Run-II has ushered in a new era in terms of energy, luminosity and discovery potential, so far there is no unambiguous signal of NP beyond the Standard Model (BSM). On the other hand,  several intriguing hints of  discrepancies between  the observed data and the SM predictions have been reported  by the $B$-physics experiments, i.e., Belle, BaBaR and LHCb, in the last few years. These discrepancies are mainly in the form of lepton flavor universality violations in semileptonic $B$ decays associated with the charged current (CC) $b \to c \ell \bar{\nu}$~\cite{Lees:2012xj,Lees:2013uzd,Aaij:2015yra,Sato:2016svk,Huschle:2015rga,Hirose:2016wfn,Aaij:2017uff,Abdesselam:2019dgh,Aaij:2017deq} and neutral current (NC) $b \to s \ell^+ \ell^-$~\cite{Aaij:2013aln,Aaij:2013qta,Aaij:2014pli,Aaij:2015esa,Aaij:2015oid,Aaij:2020nrf,Aaij:2014ora,Aaij:2017vbb,Aaij:2019wad,Abdesselam:2019lab,Abdesselam:2019wac} transitions. In the absence of any much anticipated direct NP signal at the LHC experiment, these tantalizing   hints of LFU violating   observables  play a crucial role in exploring the BSM physics and thus have attracted immense attention in the last few years.

Sizeable deviations have been observed by three different  experiments in the LFU observables of the charged-current channels, which are characterized as the ratios of branching fractions
\begin{eqnarray}
R_{D^{(*)}}\equiv \frac{{\rm Br}(B \to D^{(*)} \tau \bar \nu)}{{\rm Br}(B \to D^{(*)} \ell \bar \nu)},
\end{eqnarray}  
with $\ell=e$ or $\mu$ and
\begin{eqnarray}
R_{J/\psi}\equiv \frac{{\rm Br}(B_c \to J/\psi \tau \bar \nu)}{{\rm Br}(B_c \to J/\psi \mu \bar \nu)}.
\end{eqnarray} 
These observables are considered  as the clean probes of NP as the hadronic uncertainties inherent in individual branching fraction predictions canceled out to a large extent. The present world averages of $R_{D^{(*)}}$ measurements, performed by the Heavy Flavor Averaging Group (HFLAV)~\cite{Amhis:2019ckw} 
\begin{eqnarray}
R_D^{\rm exp}=0.340 \pm 0.027\pm 0.013, ~~~~~~~R_{D^*}^{\rm exp}=0.295 \pm 0.011\pm0.008, 
\end{eqnarray} 
have $3.1 \sigma$ deviations (including their correlation of $-0.38$) from the corresponding SM predictions
$R_D^{\rm SM}=0.299 \pm 0.003 ~(1.4 \sigma)$  and $R_{D^*}^{SM}=0.258 \pm 0.005~ (2.5 \sigma)$~\cite{Amhis:2019ckw}. In the same line, the measured ratio  $R_{J/\psi} = 0.71 \pm 0.17 \pm 0.18$~\cite{Aaij:2017tyk} also has $1.7\sigma$ deviation from its SM prediction, $R_{J/\psi}^{\rm SM}=  0.289 \pm 0.010$~\cite{Dutta:2017xmj}. Moreover, the recent measurement of the
longitudinal polarization of $D^*$ meson in $B^0  \to D^{*- }\tau^+ \bar \nu$ by Belle collaboration, $F_L^{D^*}= 0.60 \pm  0.08 \pm 0.04 $~\cite{Abdesselam:2019wbt},  also differs from its SM value $0.46\pm 0.04$~\cite{Alok:2016qyh} by $1.6 \sigma$. These deviations primarily hint towards the possible interplay of NP in $b \to c \tau \bar \nu$ decay channels. Recently, these anomalies have been studied in various model independent techniques~\cite{Bhattacharya:2018kig,Hu:2018veh,Alok:2019uqc,Asadi:2019xrc,Murgui:2019czp,Blanke:2019qrx,Shi:2019gxi,Becirevic:2019tpx,Sahoo:2019hbu,Cheung:2020sbq,Kumbhakar:2020jdz}.

The LFU violation observables in the neutral current sector are associated with $b \to s \ell^+ \ell^-$ transition and are described as 
\begin{eqnarray}
R_{K^{(*)}}\equiv \frac{{\rm Br}(B \to K^{(*)} \mu^+  \mu^-)}{{\rm Br}(B \to K^{(*)} e^+e^-)},
\end{eqnarray}
which also show around $(2-2.5)\sigma$ deviation from their SM values~\cite{Aaij:2014ora,Aaij:2017vbb,Aaij:2019wad,Abdesselam:2019lab,Abdesselam:2019wac}. In addition, the measured values of the branching fraction of $B_s\to \phi\mu^+\mu^-$~\cite{Aaij:2013aln,Aaij:2015esa} and the angular observable $P'_5$ in $B\to K^*\mu^+\mu^-$ decay~\cite{Aaij:2013qta,Aaij:2015oid,Aaij:2020nrf} differ from their SM predictions at the level of $(3-3.5)\sigma$. Assuming NP contributes  only in $b\to s\mu^+\mu^-$ transition, it has been shown that the allowed NP solutions can  be described in the form of vector and axial-vector operators. Recent global fit studies for this sector can be found in Refs.~\cite{Alok:2019ufo,Alguero:2019ptt,Ciuchini:2019usw,Kowalska:2019ley,Arbey:2019duh,Aebischer:2019mlg}.

These observed hints of LFU violation  have triggered a large number of detailed  phenomenological studies trying to ascertain the nature of plausible NP  explanation.  As the $b \to c \ell \bar \nu$ CC  transitions  occur at the tree-level,  while the NC transitions $b \to s \ell^+ \ell^-$ appear one-loop level, the anomalies associated with these transitions probe essentially different NP scales. Therefore, most of the theoretical studies in the literature have attempted to   address either NC or CC oddities, but not both on the same footing. There exists only few scenarios which can simultaneously accommodate both these anomalies and   Leptoquark (LQ) model is one such possible framework \cite{Bauer:2015knc, Becirevic:2016yqi,Fajfer:2015ycq, Sahoo:2016pet, Chen:2017hir, Crivellin:2018yvo,Kumar:2018kmr,Cornella:2019hct}.  The existence  of  LQs at low energy  is predicted in many extensions of the standard model such as Grand unified theory (GUT) \cite{Georgi:1974sy,Georgi:1974my,Fritzsch:1974nn, Langacker:1980js},
Pati-Salam model \cite{Pati:1974yy, Pati:1973uk, Pati:1973rp},  technicolor \cite{Shanker:1981mj, Schrempp:1984nj, Kaplan:1991dc}, composite model \cite{Gripaios:2009dq}  etc.

Concerning the recent flavor anomalies, the $U_1$ vector LQ which transforms as $(3,1,2/3)$ under the SM gauge group $SU(3)_C \times SU(2)_L \times U(1)_Y$ is known to  successfully elucidate them. Therefore, in this work, we would like to investigate in detail the effect of  $U_1(3,1,2/3)$ LQ on another class of semileptonic rare  $B$ decays, mediated through $b \to u \tau\bar \nu$ transitions.  We would like to emphasize here  that,  for $b \to c$ anomalies, it is customarily assumed that the NP is coupled only to the third generation leptons rather than the first two generations, i.e., in the $b \to c \tau \bar \nu$ processes. Hence, it is natural to expect that  the same class of NP might also show up in the rare processes involving   $b \to u \tau \bar \nu$ transition. Furthermore, as these CC transitions are doubly Cabibbo suppressed, the NP contributions could be significant enough leading to sizeable effects in some of the observables.  Recently, some groups have addressed different NP effects on various decays mediated by $b\to u$ transition~\cite{Rajeev:2018txm,Colangelo:2019axi,Sahoo:2020wnk,Colangelo:2020vhu}.

The outline of the paper is as follows. In Section II, we discuss briefly the relevant effective Hamiltonian describing the  semileptonic transitions $b 
\to (c, u) \ell \bar \nu$ and $b \to s \ell^+ \ell^-$. The NP contributions arising from the exchange of vector LQ $U_1$ is presented in Section III. Section IV contains the discussion about our numerical fit technique and the constraints  obtained on the NP parameters. The implications of vector LQ on various decay observables of $b \to u \tau \bar \nu$ processes are presented in section V and our conclusions are summarized in Section VI.

\section{Effective Hamiltonians for $b\to c (u)\tau\bar{\nu}$ and $b\to s \ell^+ \ell^-$}
The most general effective Hamiltonian for the charged current transition $b\to c\tau\bar{\nu}$ can be written as 
\begin{equation}
\mathcal{H}^{b\to c}_{\rm eff} = \frac{4G_F}{\sqrt{2}}V_{cb} \left[(1+C^{b\to c}_{V_L})\mathcal{O}_{V_L} + C^{b\to c}_{V_R}\mathcal{O}_{V_R}+C^{b\to c}_{S_L}\mathcal{O}_{S_L}+C^{b\to c}_{S_R}\mathcal{O}_{S_R}+ C^{b\to c}_T\mathcal{O}_T\right],
\end{equation}
where $G_F$ is the Fermi constant and $V_{cb} = (42.2\pm 0.08)\times 10^{-3}$~\cite{Tanabashi:2018oca} is the Cabibbo-Kobayashi-Maskawa (CKM) matrix element.  Here we assume that the neutrino is left-chiral. The operator  $\mathcal{O}_{V_L}$ is the SM four-fermion interaction which has the usual $(V-A)\times (V-A)$ structure, whereas $\mathcal{O}_{V_R, S_L, S_R, T}$ are the new operators which arise only in beyond the SM scenarios. The $C^{b\to c}_i$  ($i=V_L, V_R, S_L, S_R, T)$ are the corresponding NP Wilson coefficients (WCs). The explicit forms of the SM as well as NP operators are
\begin{eqnarray}
&&\mathcal{O}_{V_L} =(\bar c \gamma_{\mu} P_L b)(\bar \tau \gamma^{\mu}  P_L \nu) \ , \quad  
\mathcal{O}_{V_R}=(\bar c \gamma_{\mu}  P_R b)(\bar \tau \gamma^{\mu}  P_L \nu) \nonumber \\
&&\mathcal{O}_{S_L}=(\bar c P_L b)(\bar \tau P_L \nu ), \quad
\mathcal{O}_{S_R}=(\bar c P_R b)(\bar \tau P_L \nu),  \quad    
\mathcal{O}_T=(\bar c \sigma_{\mu \nu}P_L b)(\bar \tau \sigma^{\mu \nu} P_L \nu) \ ,
\label{opsbc} 
\end{eqnarray}
where $P_{L,R} = (1\mp \gamma_5)/2$ are the chiral projection operators. %{\color{red}A global fit to the current $b\to c\tau\bar{\nu}$ data lead to only $\mathcal{O}_{V_L}$ solution for the case of one operator at a time~\cite{Alok:2019uqc}.} 

Analogously, the effective Hamiltonian for $b\to u\tau\bar{\nu}$ transition can be expressed as
\begin{equation}
\mathcal{H}^{b\to u}_{\rm eff} = \frac{4G_F}{\sqrt{2}}V_{ub} \left[(1+C^{b\to u}_{V_L})\mathbb{O}_{V_L} + C^{b\to u}_{V_R}\mathbb{O}_{V_R}+C^{b\to u}_{S_L}\mathbb{O}_{S_L}+C^{b\to u}_{S_R}\mathbb{O}_{S_R}+ C^{b\to u}_T\mathbb{O}_T\right],
\end{equation}
where $V_{ub} = (3.94\pm 0.36)\times 10^{-3}$~\cite{Tanabashi:2018oca} is the relevant CKM matrix element. The five  operators $\mathbb{O}_i$ for this transition take the same structure as in Eq.~(\ref{opsbc}) with $c$ quark being replaced by an $u$ quark. The $C_i^{b\to u}$ are the NP WCs for $b\to u \tau\bar{\nu}$ transition.

The SM effective Hamiltonian for the FCNC decays mediated by the quark level transition $b\to s\ell^+\ell^-$ is
\begin{align} \nonumber
\mathcal{H}_{\rm SM} &= − \frac{4 G_F}{\sqrt{2} \pi} V_{tb} V_{ts}^*  \bigg[ \sum_{i=1}^{6} C_i(\mu) \mathcal{O}_i(\mu) + C_7 \frac{e}{16 \pi^2} [\overline{s} \sigma_{\mu \nu}(m_s P_L  + 
m_b P_R)b]F^{\mu \nu} \\ \nonumber 
& + C_9 \frac{\alpha_{em}}{4 \pi}(\overline{s} \gamma^{\mu} P_L b)(\overline{\ell} \gamma_{\mu} \ell) + C_{10} \frac{\alpha_{em}}{4 \pi} (\overline{s} \gamma^{\mu} P_L b)(\overline{\ell} \gamma_{\mu} \gamma_5 \ell) \bigg],
\end{align}
where $V_{tb}$ and $V_{ts}$ are the CKM matrix elements and $\alpha_{em}$ is the fine structure constant. The effect of the operators $\mathcal{O}_i,\,i=1-6,8 $ can be embedded in the redefined effective WCs as $C_7(\mu)\rightarrow C^{\rm eff}_7(\mu,q^2)$ and $C_9(\mu)\rightarrow C^{\rm eff}_9(\mu,q^2)$.

We consider the addition of vector and axial-vector NP operators to the SM effective Hamiltonian of $b \rightarrow s \mu^+ \mu^-$. Consequently, the effective Hamiltonian  takes the form
\begin{align}
\mathcal{H}^{b\to s}_{\rm eff} &= \mathcal{H}_{\rm SM} + \mathcal{H}_{\rm VA} ,
\end{align}
where $\mathcal{H}_{\rm VA}$ is expressed as
\begin{eqnarray}
\mathcal{H}_{\rm VA} &=& \frac{\alpha_{em}\,G_F}{\sqrt{2} \pi} V_{tb}  V_{ts}^* \bigg[C_9^{\rm NP}(\overline{s} \gamma^{\mu} P_L b)(\overline{\mu} \gamma_{\mu} \mu) + C_{10}^{\rm NP} (\overline{s} \gamma^{\mu} P_L b)(\overline{\mu} \gamma_{\mu} \gamma_{5} \mu) \\ \nonumber
& +& C_9^{\prime \rm NP}(\overline{s} \gamma^{\mu} P_R b)(\overline{\mu} \gamma_{\mu} \mu) + C_{10}^{\prime \rm NP} (\overline{s} \gamma^{\mu} P_R b)(\overline{\mu} \gamma_{\mu} \gamma_{5} \mu)\bigg]. 
\end{eqnarray}
Here $C^{\rm NP}_{9, 10}$ and $C^{\prime \rm NP}_{9, 10}$ are the NP WCs. Considering one operator at a time, it has been shown  in Ref.~\cite{Alok:2019ufo}, that there are only three possible NP solutions: (I) $C^{\rm NP}_9= -1.09\pm 0.18$, (II) $C^{\rm NP}_9 = -C^{\rm NP}_{10} = -0.53\pm 0.09$ and (III) $C^{\rm NP}_9 = -C^{\prime \rm NP}_9 = -1.12\pm 0.17$, which can account for present data in this sector. 
 
\section{NP effects in vector LQ model}
We now consider the effect of vector LQ $U_1(3,1,2/3)$ on these decay processes. This LQ can  explain the anomalies in both $b\to c \tau\bar{\nu}$ and $b\to s \mu^+ \mu^-$ transitions~\cite{Kumar:2018kmr,Cornella:2019hct}. The interaction Lagrangian of $U_1$ LQ with the SM fermions can be written as
\begin{equation}
\mathcal{L}_{\rm LQ}^{U_1} = h^{ij}_L \bar{Q}_{iL}\gamma_{\mu} L_{jL} U_1^{\mu}+ h^{ij}_R \bar{d}_{iR}\gamma_{\mu}l_{jR} U^{\mu}_1 + h.c. , \label{Lan}
\end{equation}
where $h^{ij}_{L,R}$ are the couplings of $U_1$  to quark and lepton pairs, with $i, j$ being their respective generation indices. Here $Q_{L}$ ($L_{L}$) is the SM left-handed quark (lepton) doublet whereas $d_R$ ($l_R$) is the right-handed down quark (lepton) singlet. The Lagrangian in Eq.~\ref{Lan} is written in the weak basis of the fermionic fields. Transforming into the mass  basis and using the Fierz identities, we can obtain relations between the LQ couplings and the NP WCs of $b\to c\tau\bar{\nu}$ transitions. Thus, one can obtain the following relations
\begin{eqnarray}
C^{b\to c}_{V_L} &= & \frac{1}{2\sqrt{2}G_F V_{cb}}\sum_{k=1}^{3} V_{k3}\frac{h^{23}_L\, h^{k3*}_L}{M^2_{U_1}} \simeq \frac{1}{2\sqrt{2}G_F V_{cb}} V_{33}\frac{h^{23}_L\, h^{33*}_L}{M^2_{U_1}}, \nonumber\\
C^{b\to c}_{S_R} &= & -\frac{1}{2\sqrt{2}G_F V_{cb}}\sum_{k=1}^{3} V_{k3}\frac{2 h^{23}_L\, h^{k3*}_R}{M^2_{U_1}} \simeq -\frac{1}{2\sqrt{2}G_F V_{cb}} V_{33}\frac{2 h^{23}_L\, h^{33*}_R}{M^2_{U_1}},
\label{WCbc}
\end{eqnarray}
where $V_{k3}$ is the CKM matrix elements and $M_{U_1}$ is the mass of the LQ, which is assumed to be 1 TeV in this analysis. To get the final expressions, we neglect the terms containing $V_{13}$ and $V_{23}$ as they are  Cabbibo suppressed.  For $b\to u\tau\bar{\nu}$ transition,  the relations in Eq.~\ref{WCbc} can be written as
\begin{eqnarray}
C^{b\to u}_{V_L} &= & \frac{1}{2\sqrt{2}G_F V_{ub}} V_{33}\frac{h^{13}_L\, h^{33*}_L}{M^2_{U_1}}, \nonumber\\
C^{b\to u}_{S_R} &= & -\frac{1}{2\sqrt{2}G_F V_{ub}} V_{33}\frac{2 h^{13}_L\, h^{33*}_R}{M^2_{U_1}}.
\label{WCbu}
\end{eqnarray}

This LQ can also generate the interaction terms for $b\to s\mu^+\mu^-$ transition. The NP WCs in $b\to s\mu^+\mu^-$ can be expressed in terms of the LQ couplings as 
\begin{equation}
C^{\rm NP}_{9} = -C^{\rm NP}_{10} = \frac{\pi}{\sqrt{2}G_F V_{tb} V^*_{ts}\alpha_{em}}\frac{h^{22}_L\,h^{32*}_L}{M^2_{U_1}}.
\end{equation}
This particular choice is motivated from the global fit of $b\to s\mu^+\mu^-$ data. From the global fit \cite{Alok:2019ufo}, $C^{\rm NP}_9 = -C^{\rm NP}_{10}$ is the only solution which can be addressed by  $U_1$ LQ scenario.

\section{Fit Methodology and Results}  

In this section, we describe the details of our fitting procedure to determine the LQ couplings $h^{13}_L$ and $h^{33}_R$ for $b\to u\tau\bar{\nu}$ transition.  We assume the value of $h^{33}_L$ to be 1 because of the hierarchy in coupling constants of left-chiral particles in flavor basis. We also assume these couplings to be real. To obtain the values of $h^{13}_L$ and $h^{33}_R$, we perform a $\chi^2$ analysis by using the CERN minimization code {\tt MINUIT}~\cite{James:1975dr,James:1994vla}. In doing so, we use the data from $b\to c\tau\bar{\nu}$, $b\to s\mu^+\mu^-$ and $b\to u\tau\bar{\nu}$ transition processes. Thus, the total $\chi^2$ is expressed as
\begin{equation}
\chi^2_{\rm total} = \chi^2_{b\to c\tau\bar{\nu}} + \chi^2_{b\to s\mu^+\mu^-} + \chi^2_{b\to u\tau\bar{\nu}}.
\end{equation}
Below we provide the discussion about the individual $\chi^2$ function in detail.

In $b\to c\tau\bar{\nu}$ sector, we take the current data of $R_D$, $R_{D^*}$, $R_{J/\psi}$ and $F_L^{D^*}$ in our fit. We do not include measurement of the $\tau$ polarization in $B\to D^*\tau\bar{\nu}$ decay because of its large statistical uncertainty~\cite{Hirose:2016wfn}. Therefore, the $\chi^2$ function for this sector looks as follows
\begin{equation}
\chi^2_{b\to c\tau\bar{\nu}} =\sum_{R_D, R_{D^*}, R_{J/\psi}, F^{D^*}_L}\left(O^{\rm th}(C^{b\to c}_i)-O^{\rm exp}\right) \mathcal{C}^{-1} \left(O^{\rm th}(C^{b\to c}_i)-O^{\rm exp}\right),
\label{chi2}
\end{equation}
where $O^{\rm th}(C^{b\to c}_i)$ are NP predictions of each observable and $O^{\rm exp}$ are the corresponding experimental central values. Here $\mathcal{C}$ denotes the covariance matrix which includes both theory and experimental correlations. We also include the constraint from the branching fraction of $B_c\to \tau\bar{\nu}$. We set the upper limit of this quantity to be $30\%$ which is calculated from the lifetime of $B_c$ meson~\cite{Alonso:2016oyd}.

In the context of $U_1$ LQ, the NP WCs in $b\to s\mu^+\mu^-$ transition  are related as $C^{\rm NP}_9 = -C^{\rm NP}_{10} = -0.53\pm 0.09$~\cite{Alok:2019ufo}. Hence, we can use this result to constrain the LQ couplings. For this sector, we define the $\chi^2$ function as
\begin{equation}
\chi^2_{b\to s\mu^+\mu^-} = \left(\frac{C^{\rm NP}_9- (-0.53)}{0.09}\right)^2.
\end{equation} 

In $b\to u\tau\bar{\nu}$ transition, the only measured quantity is the branching fraction of $B^+\to \tau^+\nu$ process with a value $(1.09\pm 0.24)\times 10^{-4}$ \cite{Tanabashi:2018oca}. The SM prediction for this branching ratio is $(8.80\pm 0.73)\times 10^{-5}$. Therefore, there is a tension between the measured value and the SM prediction at the level of $\sim 1\sigma$. In addition, Belle collaboration has put an upper limit on the branching fraction of $B\to \pi^-\tau^+\nu$. They obtained an upper limit of $2.5\times 10^{-4}$ at the $90\%$ C.L.~\cite{Hamer:2015jsa}. Therefore, the $\chi^2$ function for this sector can be written as
\begin{equation}
\chi^2_{b\to u\tau\bar{\nu}} = \frac{\left({\rm Br}(B^+\to \tau^+\nu)-1.09\times 10^{-4}\right)^2}{(0.24\times 10^{-4})^2+ (0.73\times 10^{-5})^2} + \frac{\left({\rm Br}(B\to \pi^-\tau^+\nu)-1.25\times 10^{-4}\right)^2}{(0.76\times 10^{-4})^2}.
\end{equation} 
In writing the $\chi^2$ term for the branching fraction of $B\to \pi^-\tau^+\nu$, we have adjusted the central value and the error such that we can get the value of upper limit at a level of $1.645\sigma$ (or $90\%$ C.L.).

\begin{figure}[htbp]
\includegraphics[width=100mm]{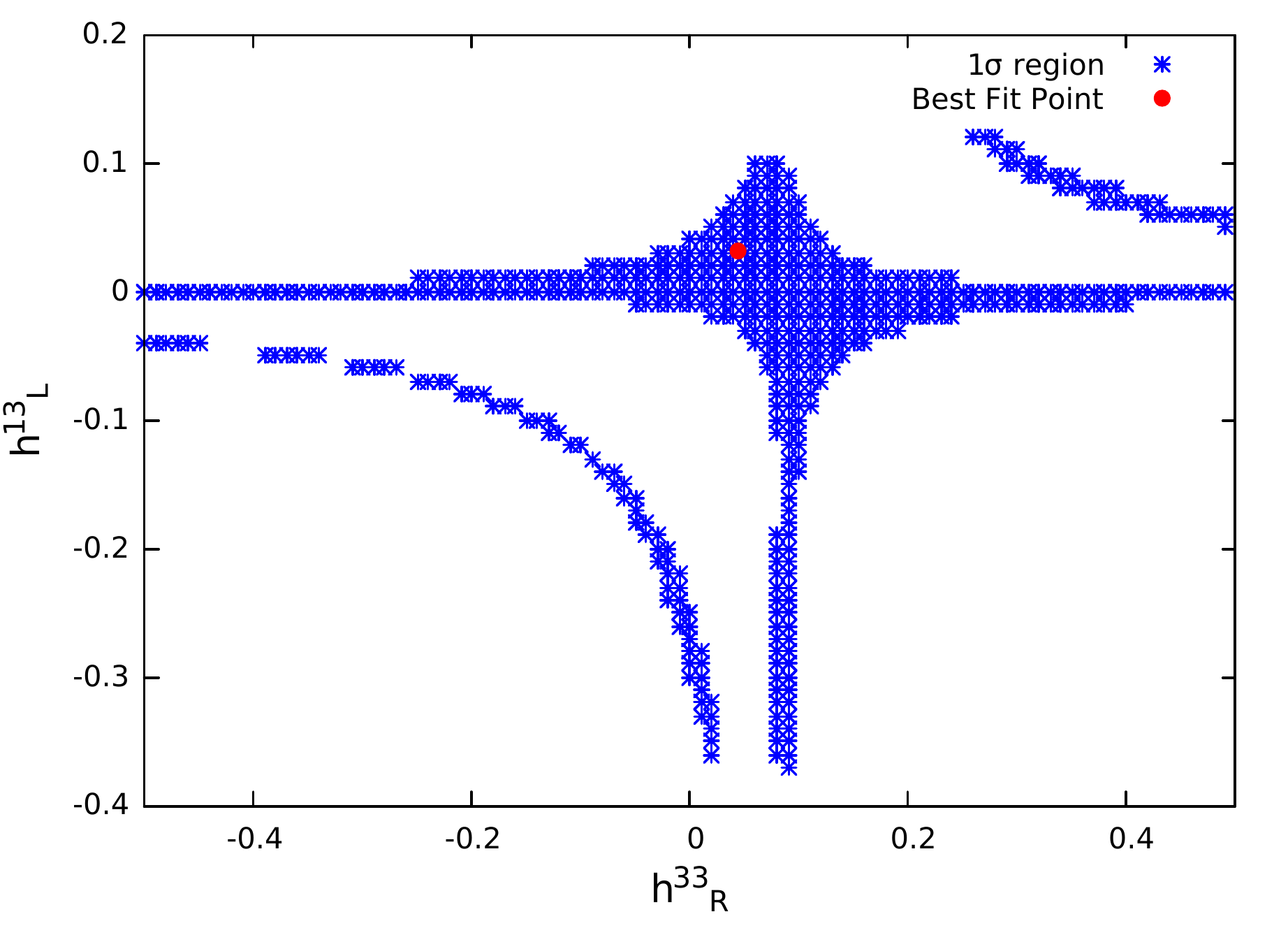}
\caption{$1\sigma$ allowed region in the $h^{13}_L-h^{33}_R$ plane, constrained by the current data from $B$ sector.}
\label{fig1}
\end{figure}

We use the {\tt Flavio} package~\cite{Straub:2018kue} to compute the observables which are taken into the fit. Minimizing the $\chi^2_{\rm total}$, we obtain the best fit values $h^{13}_L = 0.03$ and $h^{33}_R = 0.04$ of the LQ couplings for $b\to u \tau\bar{\nu}$ transition. We find the correlation between these two parameters is $\sim 0.80$. We also determine the $1\sigma$ allowed parameter space for $h^{13}_L$-$h^{33}_R$. This is shown in Fig.~\ref{fig1}. This figure shows space for NP in $b\to u$ transition allowed by current data in $B$ sector.

\section{Predictions for $b\to u \tau\bar{\nu}$ decay observables in $U_1$ LQ}
In this section, we investigate the effects of $U_1$ LQ on various decay modes mediated by $b\to u \tau\bar{\nu}$ transition. In particular, we focus on the decays $B\to \pi\tau\bar{\nu}$, $B\to (\rho, \omega)\tau\bar{\nu}$ and $B_s\to (K, K^*)\tau\bar{\nu}$. We mainly focus on the branching fraction and the lepton flavor ratio $R^{\tau/\ell}$ for each decay. In addition, we compute the standard angular observables, e.g., the forward-backward asymmetry $A_{FB}$, the $\tau$ polarization fraction $P_{\tau}$ and the longitudinal polarization fraction of vector meson $F_L$, for each decay mode. These observables  are defined as follows
\begin{equation}
\frac{d {\rm Br}}{dq^2} = \frac{d\Gamma/dq^2}{\Gamma_{\rm total}},\,  \quad R^{\tau/\ell}_{P,V}(q^2) = \frac{d\Gamma(B\to (P,V)\tau\bar{\nu})/dq^2}{d\Gamma(B\to (P,V)\ell\bar{\nu})/dq^2}, \nonumber
\end{equation}
\begin{equation}
P_{\tau}(q^2) = \frac{d\Gamma^{\lambda_{\tau}=1/2}/dq^2-d\Gamma^{\lambda_{\tau}=-1/2}/dq^2}{d\Gamma^{\lambda_{\tau}=1/2}/dq^2+d\Gamma^{\lambda_{\tau}=-1/2}/dq^2}, \quad F_L(q^2) = \frac{d\Gamma^{\lambda_{V}=0}/dq^2}{d\Gamma/dq^2}, \nonumber
\end{equation} 
\begin{equation}
A_{FB} (q^2)= \frac{1}{d\Gamma /dq^2}\left[\int^1_0\frac{d^2\Gamma }{dq^2d\cos\theta_{\tau}}d\cos\theta_{\tau}-\int^0_{-1}\frac{d^2\Gamma }{dq^2d\cos\theta_{\tau}}d\cos\theta_{\tau}\right].
\label{obs}
\end{equation} 
Here, $d\Gamma^{\lambda_{\tau}=\pm 1/2}/dq^2$ are the differential decay rates of $B\to (P,V)$ processes with the polarization of the $\tau$ lepton $\lambda_{\tau} = \pm 1/2$ whereas $d\Gamma^{\lambda_{V}=0}/dq^2$ is the decay rate of $B\to V$ decay with the polarization of vector $V$ meson $\lambda_V = 0$. In the $U_1$ LQ model, the differential decay rates for $B\rightarrow (P,V)\tau\bar{\nu}$ decays are written as ~\cite{Sakaki:2013bfa}
\begin{eqnarray}
\frac{d\Gamma(B\rightarrow P\tau\bar{\nu})}{dq^2} &=& \frac{G^2_F |V_{ub}|^2}{192\pi^3m^3_B}q^2\sqrt{\lambda_P(q^2)}\left(1-\frac{m^2_{\tau}}{q^2}\right)^2\times \nonumber\\
& & \left[ | 1+C^{b\to u}_{V_L}|^2\left[\left(1+\frac{m^2_{\tau}}{2q^2}\right)H^{s2}_{V,0}+ \frac{3m^2_{\tau}}{2q^2} H^{s2}_{V,t}\right] \right. \nonumber\\
& & \left. + \frac{3}{2}|C^{b\to u}_{S_R}|^2 H^{s2}_{S} +3 \,{\rm Re}\left[\left(1+C^{b\to u}_{V_L}\right) C^{*b\to u}_{S_R}\right]\frac{m_{\tau}}{\sqrt{q^2}}H^s_S H^s_{V,t}\right],
\end{eqnarray}
and 
\begin{eqnarray}
\frac{d\Gamma(B\rightarrow V\tau\bar{\nu})}{dq^2} &=& \frac{G^2_F |V_{ub}|^2}{192\pi^3m^3_B}q^2\sqrt{\lambda_{V}(q^2)}\left(1-\frac{m^2_{\tau}}{q^2}\right)^2 \times \nonumber \\
& & \left[\left(|1+C^{b\to u}_{V_L}|^2\right)\left[\left(1+\frac{m^2_{\tau}}{2q^2}\right)\left(H^2_{V,+}+H^2_{V,-}+H^2_{V,0}\right)+\frac{3m^2_{\tau}}{2q^2}H^2_{V,t}\right]\right.\nonumber \\
& & \left. + \frac{3}{2}|C^{b\to u}_{S_R}|^2 H^2_S + 3{\rm Re}\left[\left(1+C^{b\to u}_{V_L}\right)C^{*b\to u}_{S_R}\right]\frac{m_{\tau}}{\sqrt{q^2}}H_S H_{V,t} \right],
\label{dfV}
\end{eqnarray}
with
\begin{equation}
\lambda_{P,V}(q^2) = \left(\left(m_B-m_{P,V}\right)^2-q^2\right)\left(\left(m_B+m_{P,V}\right)^2-q^2\right).
\end{equation}
The SM decay rate for $\mu/e$ lepton can be obtained by setting the NP WCs to zero and by replacing $m_{\tau}$ with mass of $\mu/e$. The nonzero helicity amplitudes of $B\to P$ processes can be expressed in terms of the two form factors $F_{0,+}(q^2)$, characterizing $B \to P$ transitions and are given as 
\begin{equation}
H^s_{V,0}(q^2)= \sqrt{\frac{\lambda_P(q^2)}{q^2}}F_+(q^2), \quad
H^s_{V,t}(q^2) = \frac{m^2_B-m^2_P}{\sqrt{q^2}}F_0(q^2), \quad
H^s_S(q^2) = \frac{m^2_B-m^2_P}{m_b-m_u}F_0(q^2).
\end{equation}
On the other hand, the non-zero helicity amplitudes for $B\to V$ transitions can be expressed in terms of the corresponding hadronic form factors as
\begin{eqnarray}
H_{V,\pm}(q^2) & =& (m_B+m_V)A_1(q^2)\mp \frac{\sqrt{\lambda_{V}(q^2)}}{m_B+m_V}V(q^2), \nonumber\\
H_{V,0}(q^2) &=& \frac{m_B+m_V}{2m_{V}\sqrt{q^2}}\left[-(m^2_B-m^2_{V}-q^2)A_1(q^2)+\frac{\lambda_{V}(q^2)}{(m_B+m_{V})}A_2(q^2)\right], \nonumber \\
H_{V,t}(q^2) &=& -\sqrt{\frac{\lambda_{V}(q^2)}{q^2}}A_0(q^2), \nonumber \\
H_S(q^2) &=& -\frac{\sqrt{\lambda_{V}(q^2)}}{m_b+m_u}A_0(q^2).
\end{eqnarray}
All these form-factors are calculated using different techniques for different decay modes. We will discuss them individually for each case in the following subsections.
The decay distributions for the $\tau$ lepton polarizations $\lambda = \pm 1/2$ in $B\rightarrow P\,\tau\,\bar{\nu}$ decay are given by 
\begin{eqnarray}
\frac{d\Gamma^{\lambda_{\tau}=1/2}(B\rightarrow P\tau\bar{\nu})}{dq^2} &=& \frac{G^2_F |V_{ub}|^2}{192\pi^3m^3_B}q^2\sqrt{\lambda_P(q^2)}\left(1-\frac{m^2_{\tau}}{q^2}\right)^2\times \nonumber\\
& & \left[\frac{1}{2}| 1+C^{b\to u}_{V_L}|^2 \frac{m^2_{\tau}}{q^2} \left(H^{s2}_{V,0}+ 3H^{s2}_{V,t}\right)+  \frac{3}{2}| C^{b\to u}_{S_R}|^2 H^{s2}_{S} \right. \nonumber \\
& & \left. +3 \,{\rm Re}\left[\left(1+C^{b\to u}_{V_L}\right)C^{*b\to u}_{S_R}\right]\frac{m_{\tau}}{\sqrt{q^2}}H^s_S H^s_{V,t}\right],\nonumber \\
\frac{d\Gamma^{\lambda_{\tau}=-1/2}(B\rightarrow P\tau\bar{\nu})}{dq^2} &=& \frac{G^2_F |V_{ub}|^2}{192\pi^3m^3_B}q^2\sqrt{\lambda_P(q^2)}\left(1-\frac{m^2_{\tau}}{q^2}\right)^2\times   | 1+C^{b\to u}_{V_L}|^2  H^{s2}_{V,0} .
\end{eqnarray}
These distributions for $B\to V\tau\bar{\nu}$ decays are expressed as follows
\begin{eqnarray}
\frac{d\Gamma^{\lambda_{\tau}=1/2}(B\to V\tau\bar{\nu})}{dq^2} &=& \frac{G^2_F |V_{ub}|^2}{192\pi^3m^3_B}q^2\sqrt{\lambda_{V}(q^2)}\left(1-\frac{m^2_{\tau}}{q^2}\right)^2 \times \nonumber\\
& & \left[\frac{1}{2}\left( |1+C^{b\to u}_{V_L}|^2\right)\frac{m^2_{\tau}}{q^2} \left(H^2_{V,+}+H^2_{V,-}+H^2_{V,0}+3H^2_{V,t}\right)\right. \nonumber \\
& & \left. +\frac{3}{2}|C^{b\to u}_{S_R}|^2 H^2_{S} +3 {\rm Re}\left[\left(1+C^{b\to u}_{V_L}\right)\left(C^{*b\to u}_{S_R}\right)\right]\frac{m_{\tau}}{\sqrt{q^2}}H_S H_{V,t} \right],\nonumber \\
\frac{d\Gamma^{\lambda_{\tau}=-1/2}(B\to V\tau\bar{\nu})}{dq^2} &=& \frac{G^2_F |V_{ub}|^2}{192\pi^3m^3_B}q^2\sqrt{\lambda_{V}(q^2)}\left(1-\frac{m^2_{\tau}}{q^2}\right)^2 \times \nonumber \\ 
& &\left[\left(| 1+C^{b\to u}_{V_L}|^2\right)\left(H^2_{V,+}+H^2_{V,-}+H^2_{V,0}\right) \right]
\end{eqnarray}
The decay distribution of $B\to (P,V)\tau\bar{\nu}$ transitions with respect to $q^2$ and $\theta_{\tau}$ can be written as
\begin{equation}
\frac{d^2\Gamma(B\to (P,V)\tau\bar{\nu})}{dq^2d\cos\theta} = a^{P,V}_{\theta}(q^2) + b^{P,V}_{\theta}(q^2)\cos\theta + c^{P,V}_{\theta}(q^2) \cos^2\theta.
\end{equation} 
The definition of $A_{FB}$ in Eq.~\ref{obs} leads to the forward-backward asymmetry to be 
\begin{eqnarray}
A_{FB} (q^2)= \frac{1}{(d\Gamma /dq^2)}b^{P,V}_{\theta}(q^2), 
\end{eqnarray}
where $b^{P,V}_{\theta}$ are given by
\begin{eqnarray}
b^P_{\theta}(q^2) &=& \frac{G^2_F |V_{ub}|^2}{128\pi^3m^3_B}q^2\sqrt{\lambda_P(q^2)}\left(1-\frac{m^2_{\tau}}{q^2}\right)^2\times \nonumber\\
& & \left[ | 1+C^{b\to u}_{V_L}|^2 \frac{m^2_{\tau}}{q^2} H^{s}_{V,0} H^{s}_{V,t} + {\rm Re}\left[\left(1+C^{b\to u}_{V_L}\right)C^{*b\to u}_{S_R}\right]\frac{m_{\tau}}{\sqrt{q^2}}H^s_S H^s_{V,t}\right],
\end{eqnarray}
and 
\begin{eqnarray}
b^{V}_{\theta} & = & \frac{G^2_F |V_{ub}|^2}{128\pi^3m^3_B}q^2\sqrt{\lambda_{V}(q^2)}\left(1-\frac{m^2_{\tau}}{q^2}\right)^2 \left[ \frac{1}{2}\left(|1+C^{b\to u}_{V_L}|^2 \right)\left(H^2_{V,+}-H^2_{V,-}\right) \right.\nonumber \\
& & \left. + |1+C^{b\to u}_{V_L}|^2\frac{m^2_{\tau}}{q^2}H_{V,0}H_{V,t}+ {\rm Re}\left[\left(1+C^{b\to u}_{V_L}\right)C^{*b\to u}_{S_R}\right]\frac{m_{\tau}}{\sqrt{q^2}}H_S H_{V,0} \right].
\end{eqnarray}
The differential decay rate with the longitudinally polarized $V$ meson $d\Gamma^{\lambda_{V}=0}/dq^2$ can be written as
\begin{eqnarray}
\frac{d\Gamma^{\lambda_{V}=0}}{dq^2} &=&  \frac{G^2_F |V_{ub}|^2}{192\pi^3m^3_B}q^2\sqrt{\lambda_{V}(q^2)}\left(1-\frac{m^2_{\tau}}{q^2}\right)^2 \times \nonumber \\
& & \left[|1+C^{b\to u}_{V_L}|^2 \left[\left(1+\frac{m^2_{\tau}}{2q^2}\right)H^2_{V,0}+\frac{3m^2_{\tau}}{2q^2}H^2_{V,t}\right]      \right. \nonumber\\
& &\left.  +\frac{3}{2}|C^{b\to u}_{S_R}|^2 H^2_S + 3 {\rm Re}\left[\left(1+C^{b\to u}_{V_L}\right)C^{*b\to u}_{S_R}\right]\frac{m_{\tau}}{\sqrt{q^2}} H_S H_{V,t} \right].
\end{eqnarray}
After collating all the required information about various observables, we now proceed to appraise  their values for various decay modes.
\subsection{$B\to \pi\tau\bar{\nu}$ decay:} 
\begin{figure}[htbp]
\begin{tabular}{cc}
\includegraphics[width=70mm]{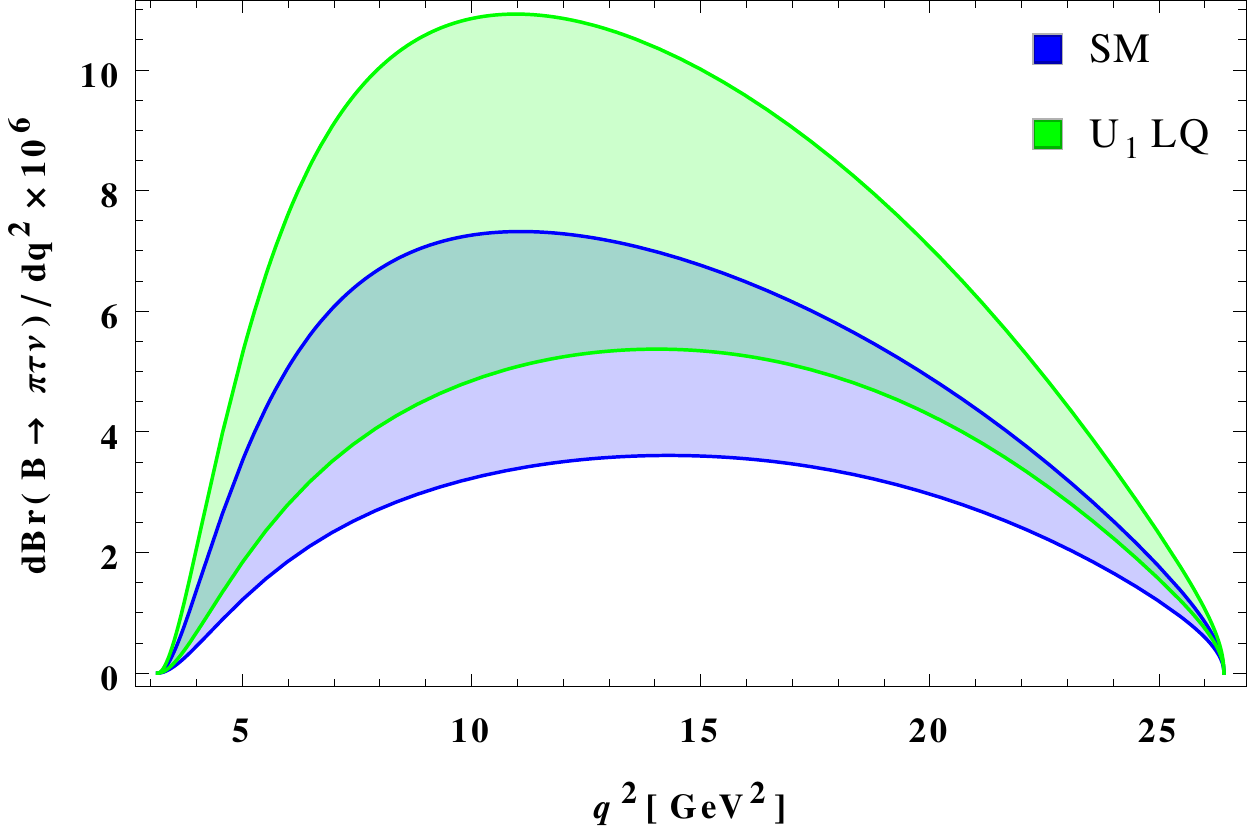} & \includegraphics[width=70mm]{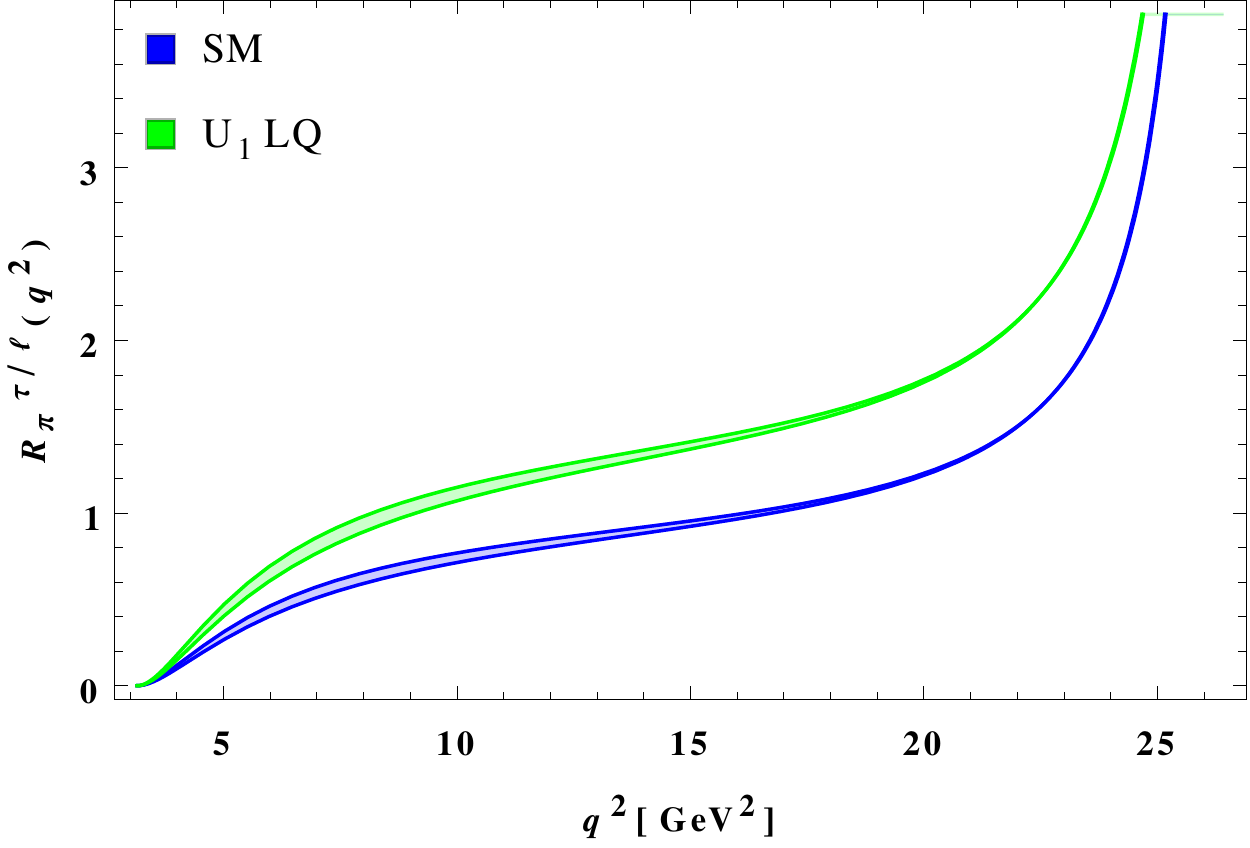} \\
\includegraphics[width=70mm]{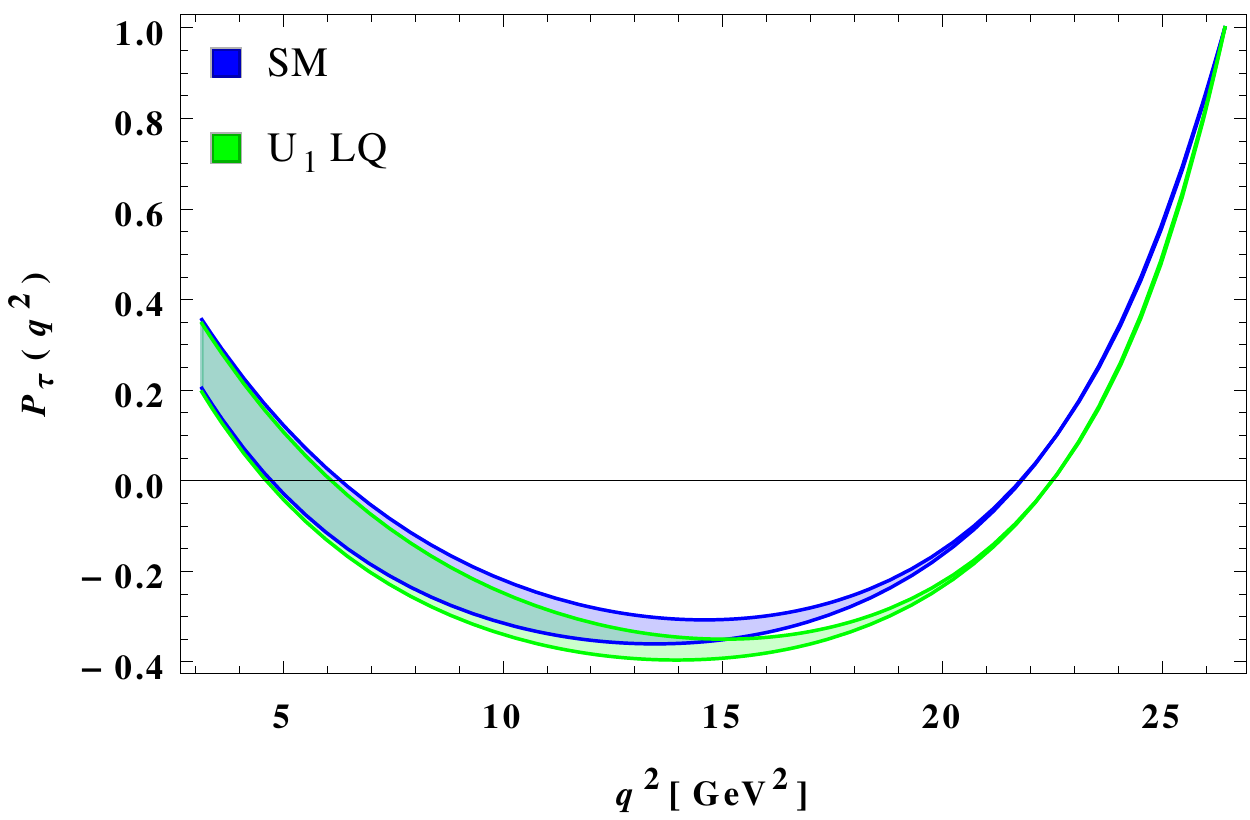} & \includegraphics[width=70mm]{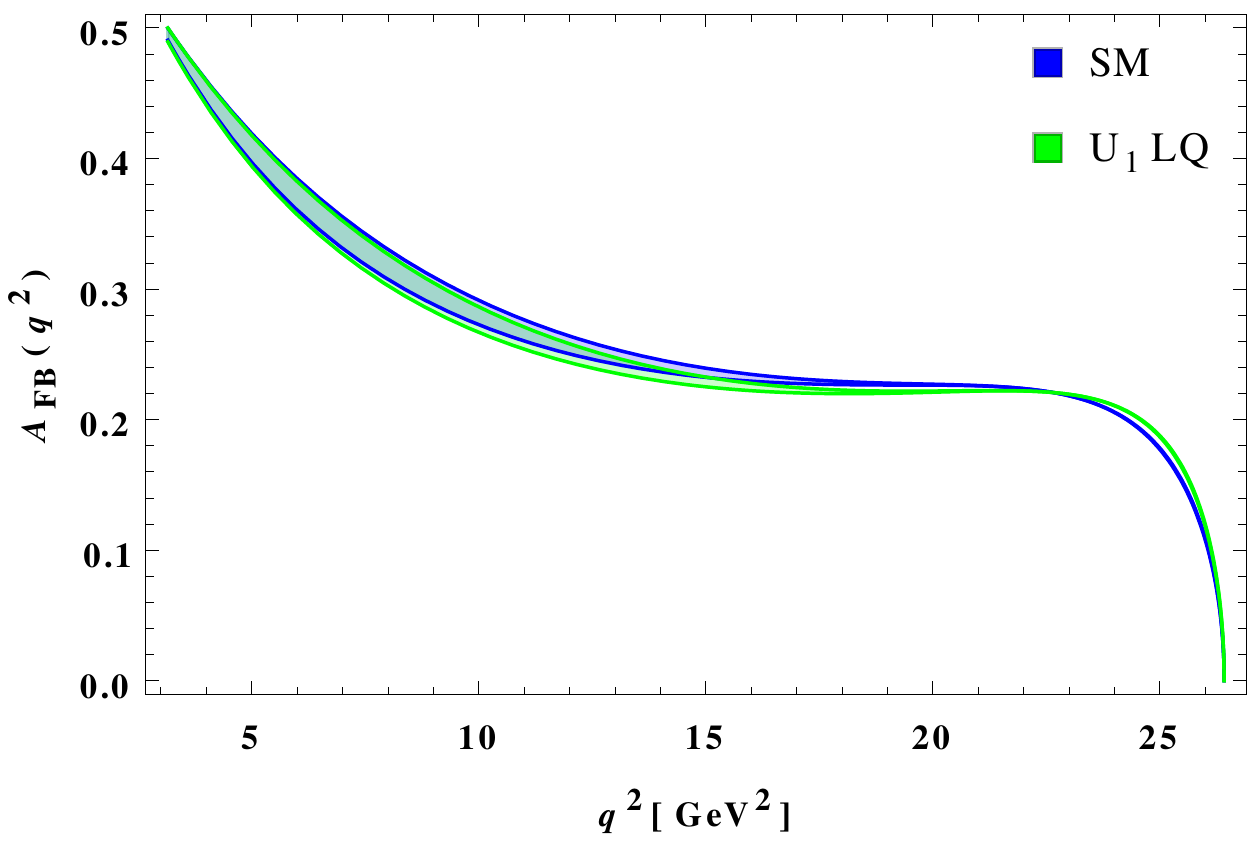} 
\end{tabular}
\caption{Variation of branching fraction (top-left panel), $R_\pi^{\tau/\ell}$ (top-right panel), $P_\tau$ (bottom-left panel) and $A_{FB}$  (bottom-right panel) with respect to $q^2$ for $B \to \pi \tau \bar \nu$ process. } 
\label{B2piFig}
\end{figure}

The form-factors $F_0$ and $F_1$ for this process are computed by lattice QCD approach, which are parametrized as follows~\cite{Lattice:2015tia}
\begin{equation}
F_+ (q^2)= \frac{1}{1-q^2/m^2_{B^*}}\sum^{N-1}_{n=0} b^+_n\left[z^n - (-1)^{n-N} \frac{n}{N}~z^N\right],\, \quad F_0(q^2) = \sum^{N-1}_{n=0} b^0_n ~z^n, 
\end{equation}
where $z (q^2) = \frac{\sqrt{t_+ -q^2}-\sqrt{t_+ -t_0}}{\sqrt{t_+ -q^2}+\sqrt{t_+ -t_0}}$,  $t_+ = (M_B+M_{\pi})^2$,  $t_0 = (M_B+M_{\pi})\left(\sqrt{M_B}-\sqrt{M_{\pi}}\right)^2$, $N = 4$ and $m_{B^*} = 5.6794(10)$ GeV. The inputs of these form-factors are given by~\cite{Lattice:2015tia}
\begin{eqnarray}
& & b^+_0 = 0.419(13), \quad b^+_1 = -0.495(54), \quad b^+_2 = -0.43(13), \quad b^+_3 = 0.22(0.31), \nonumber \\
& & b^0_0 = 0.510(19), \quad b^0_1 = -1.700(82), \quad b^0_2 = 1.53(19), \quad b^0_3 = 4.52(0.83).
\end{eqnarray}
Using these form-factors, we estimate the values of the branching fraction, $R^{\tau/\ell}_{\pi}$, $P_{\tau}$ and $A_{FB}$, for this decay mode both in the SM as well as in the $U_1$ LQ model. The variation  of these observables as a function of $q^2$ are shown in Fig.~\ref{B2piFig}. From the plots, one can notice that the impact of $U_1$ LQ on the branching fraction as well as on the lepton non-universality parameter $R^{\tau/\ell}_{\pi}$ is quite significant  whereas its effect is rather minimal on the $\tau$ polarization $P_{\tau}$ as well as on forward backward asymmetry $A_{FB}$. The predicted values of these observables both in the SM as well as in LQ scenario  are listed in Table~\ref{B2piTab}. Since the discrepancy between the SM and the LQ model predictions for the $R_{\pi}^{\tau/\ell}$ value is fairly large, it should be searched for at LHCb or Belle II experiments. 
\begin{table}[h!]
\begin{tabular}{|c|c|c|c|c|}
\hline\hline
 & ${\rm Br}(B\to \pi\tau\bar{\nu})$ & $R_{\pi}^{\tau/\ell}$ & $P_{\tau}$ & $A_{FB}$\\
 \hline
 SM &~  $(0.847\pm 0.165)\times 10^{-4}$ ~&~ $0.634\pm 0.041$ ~&~ $-0.175\pm 0.053$ ~& ~$0.262\pm 0.007$~\\
 \hline
 $U_1$ LQ & $(1.244\pm 0.242)\times 10^{-4}$ & $0.921\pm 0.057$ & $-0.236\pm 0.051$& $0.257\pm 0.008$\\
 \hline\hline
\end{tabular}
\caption{Predicted values of various observables for $B\to \pi\tau\bar{\nu}$ process, both in the SM and LQ model.}
\label{B2piTab}
\end{table}

\subsection{$B\to (\rho, \omega)\tau\bar{\nu}$ decays:}
\begin{figure}[h!]
\begin{tabular}{cc}
\includegraphics[width=70mm]{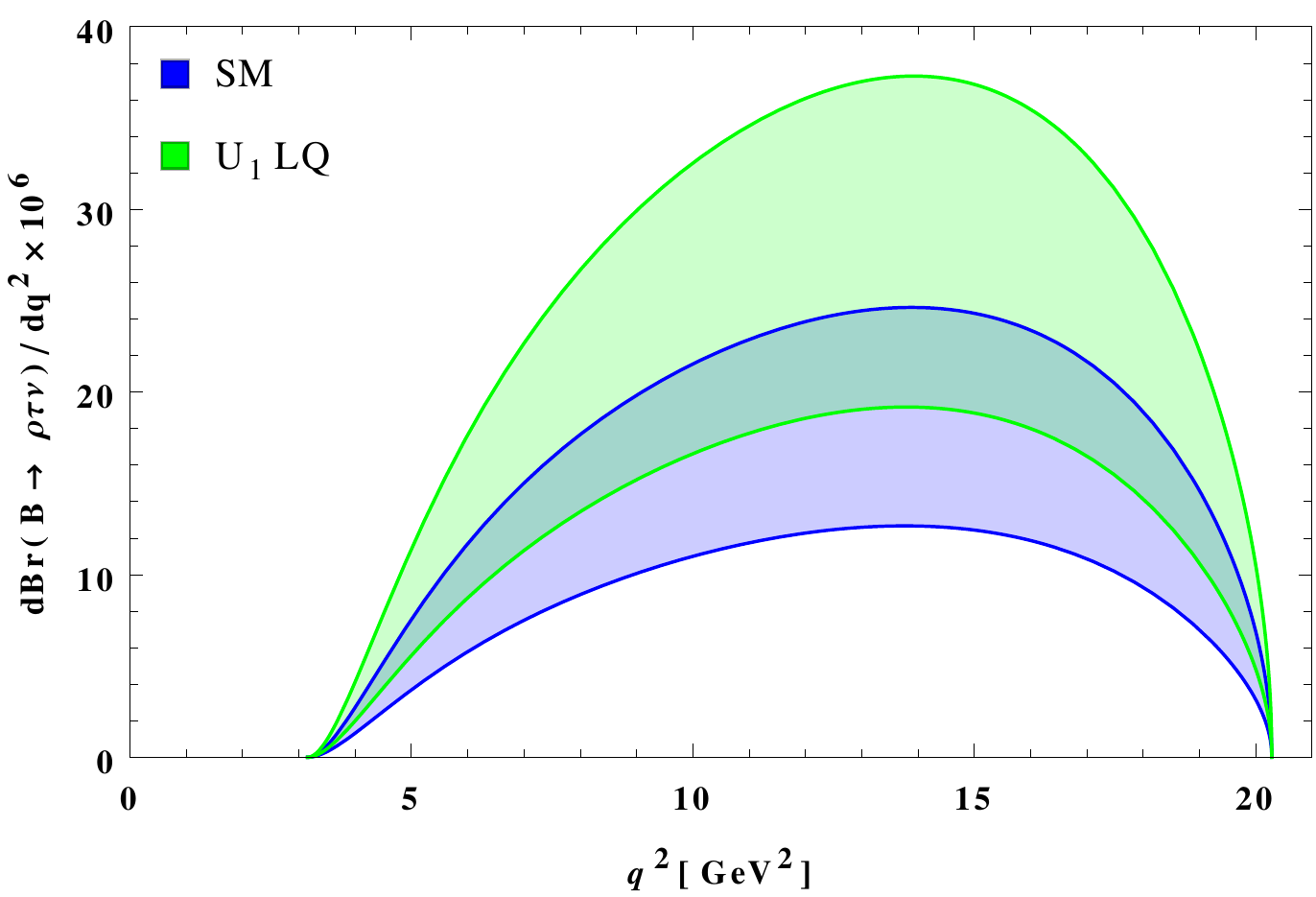} & \includegraphics[width=70mm]{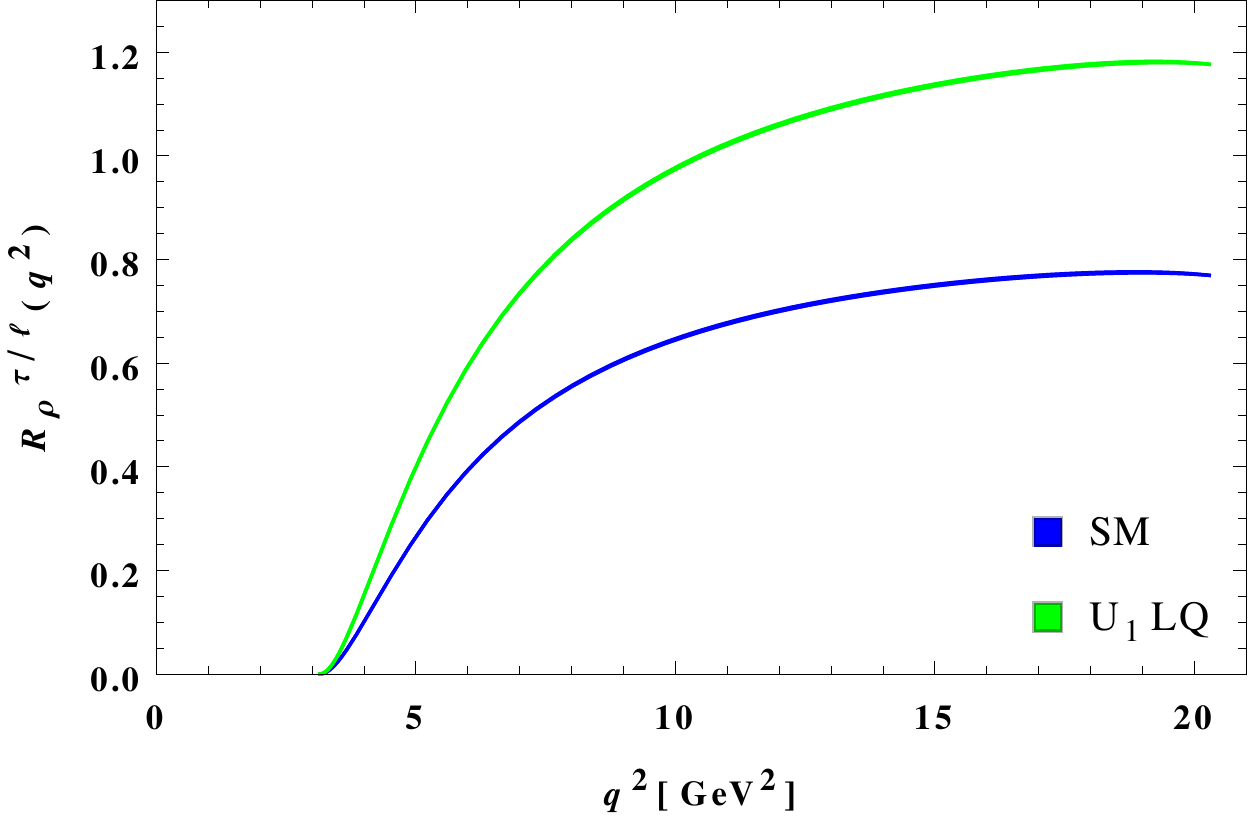} \\
\includegraphics[width=70mm]{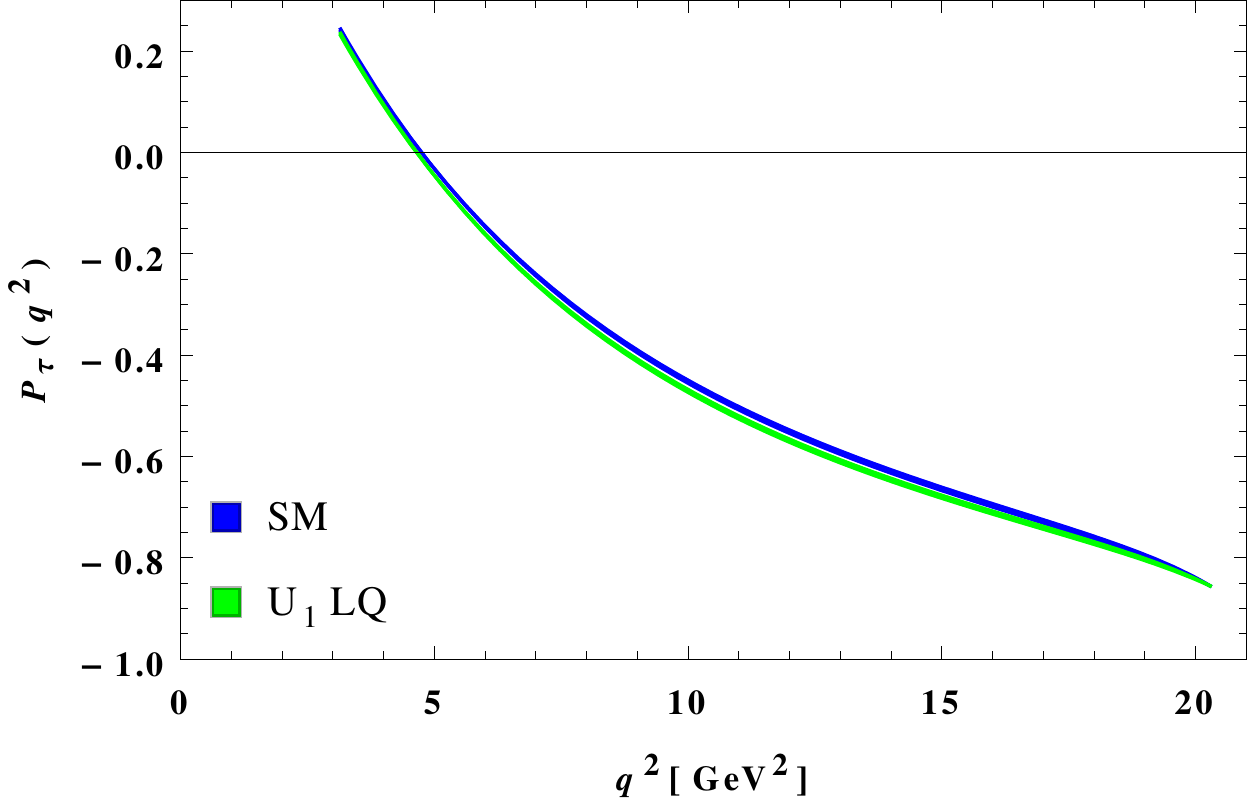} & \includegraphics[width=70mm]{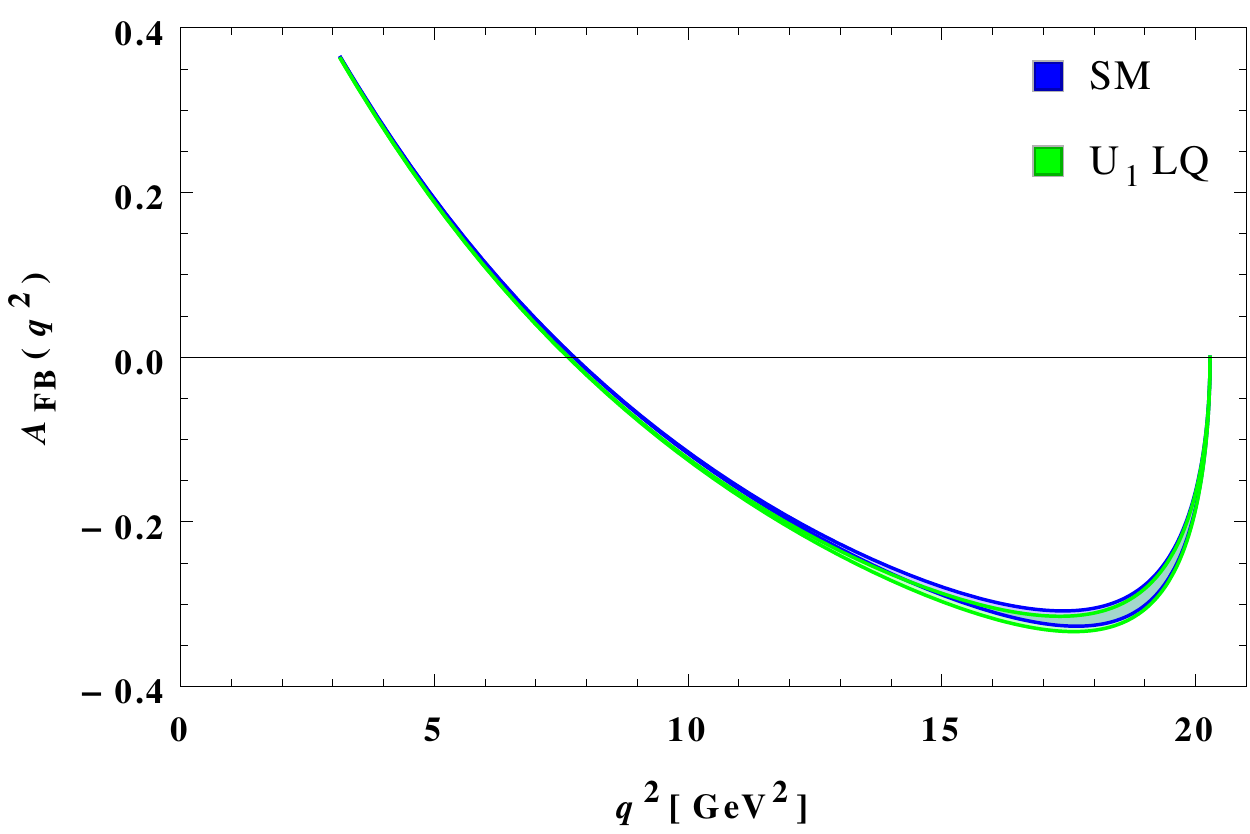} 
\end{tabular}
\includegraphics[width=70mm]{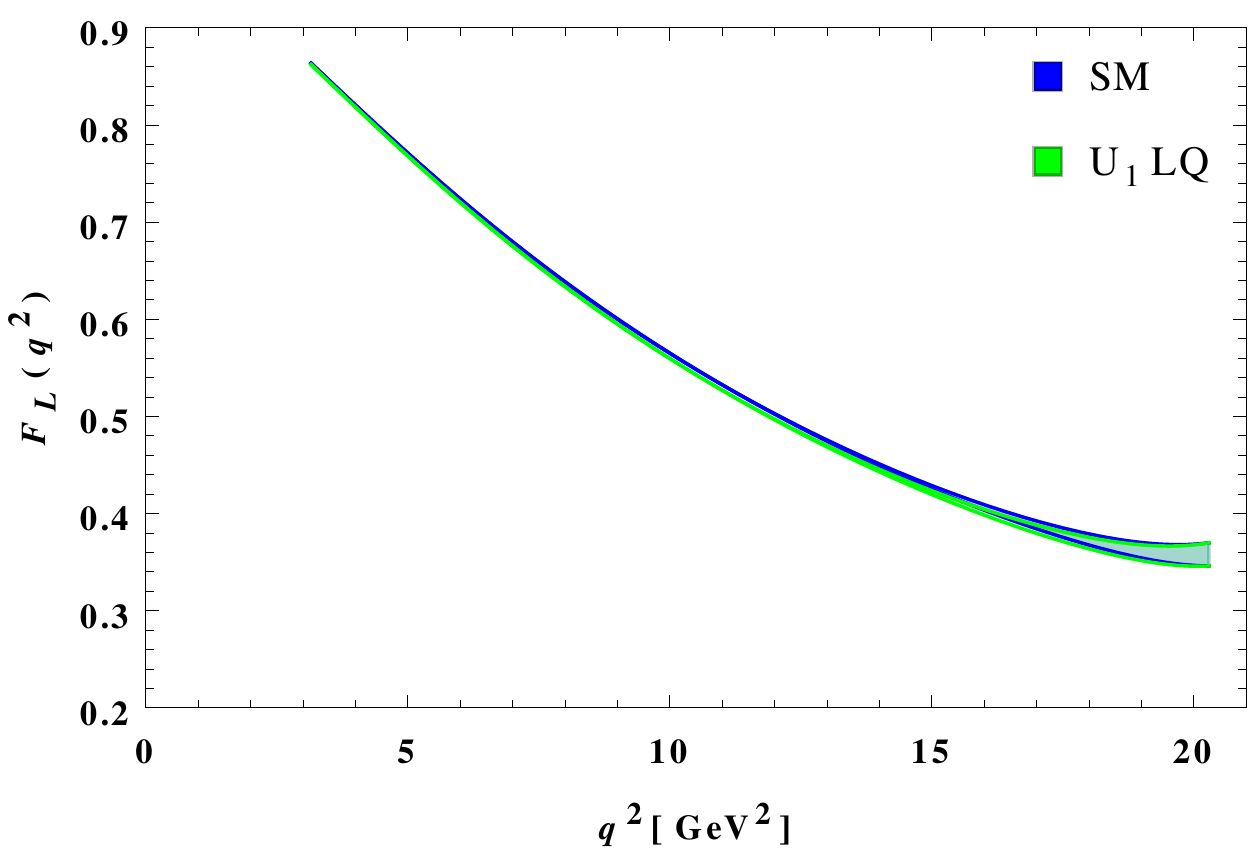} 
\caption{The $q^2$ variation of differential branching fraction, $R_\rho^{\tau/l}$, $P_\tau$, $A_{FB}$ and $F_L$  observables for $B\to \rho\tau\bar{\nu}$ process in the SM as well as in $U_1$ LQ model.}
\label{B2rhofig}
\end{figure}

\begin{figure}[h!]
\begin{tabular}{cc}
\includegraphics[width=70mm]{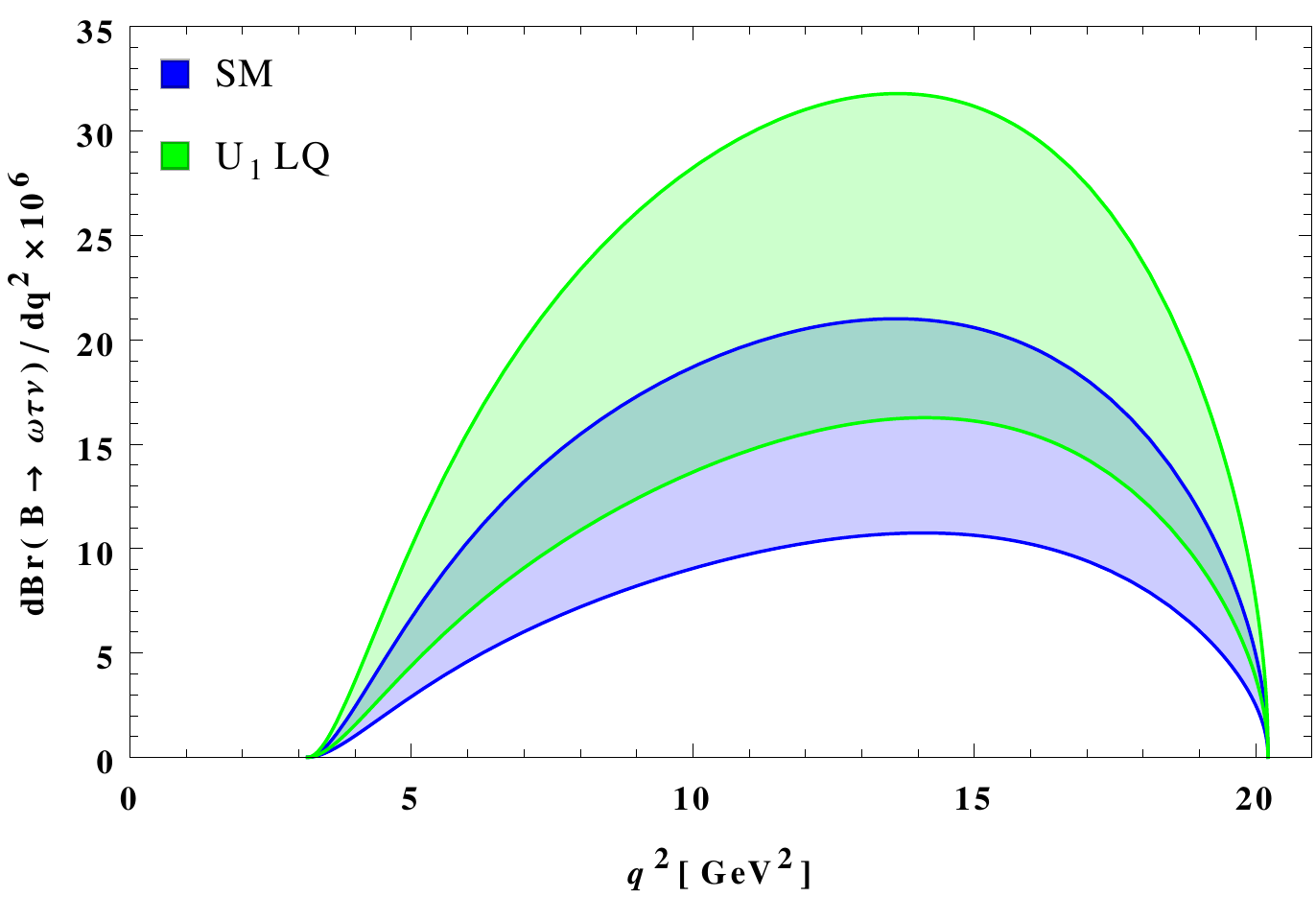} & \includegraphics[width=70mm]{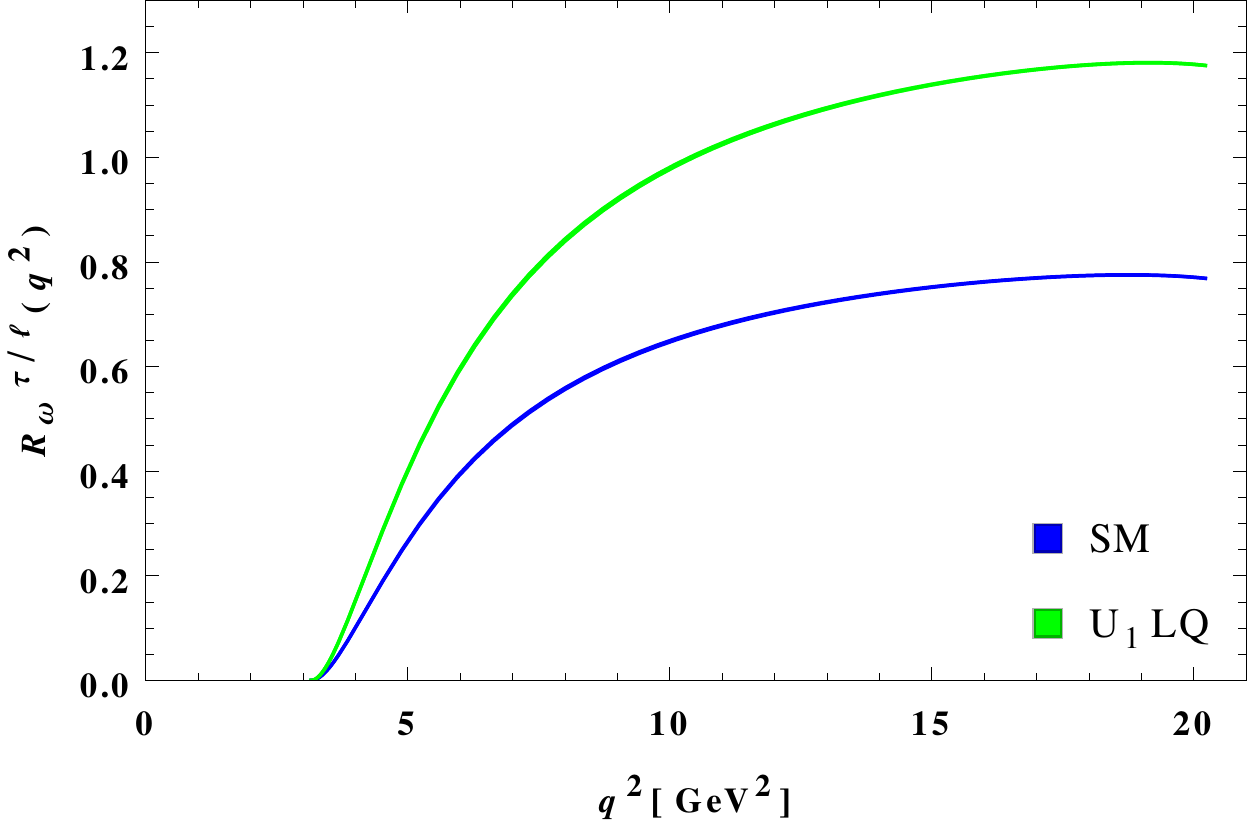} \\
\includegraphics[width=70mm]{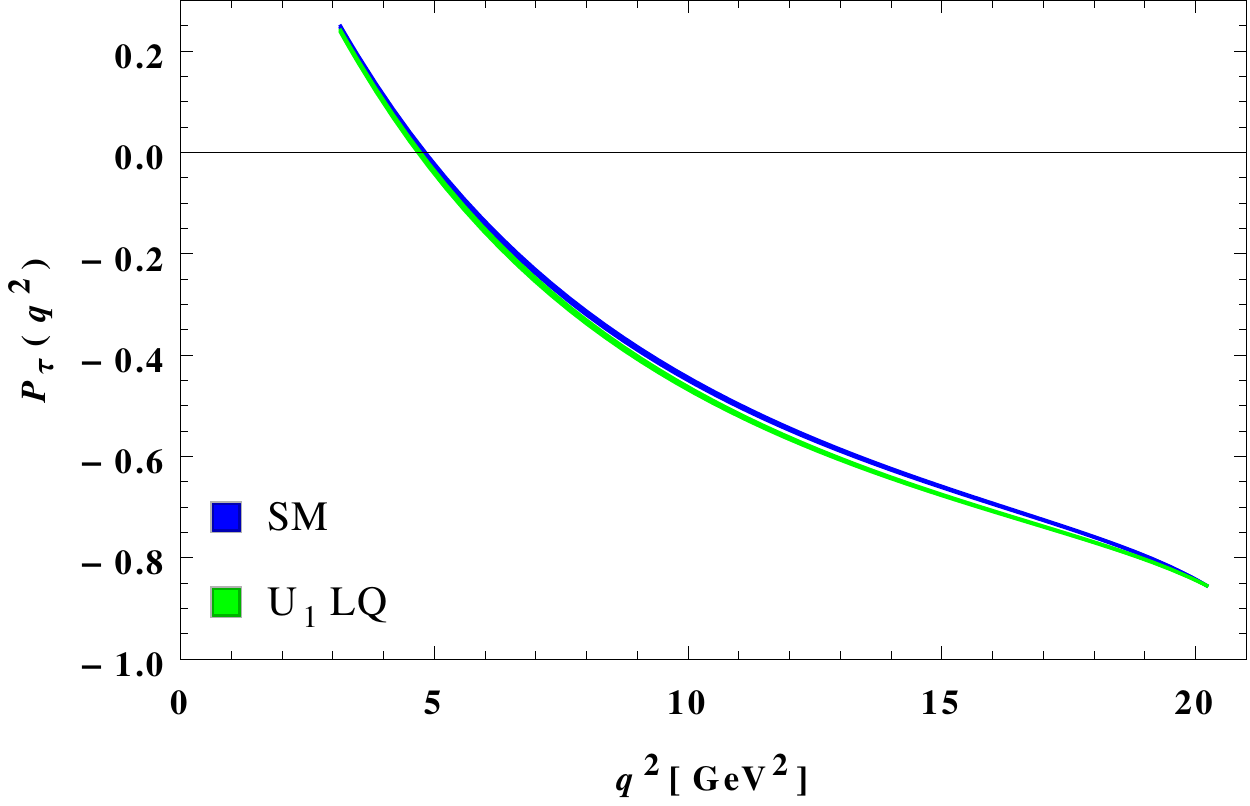} & \includegraphics[width=70mm]{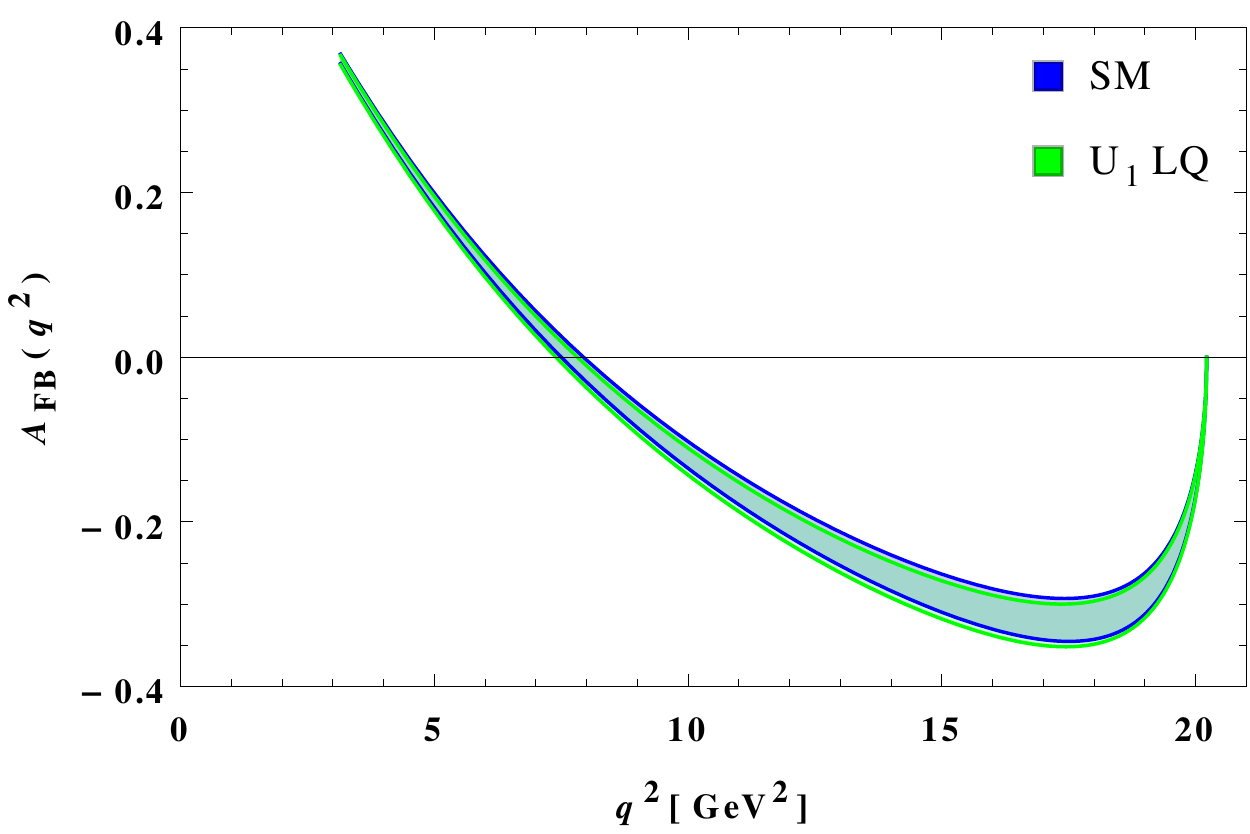} 
\end{tabular}
\includegraphics[width=70mm]{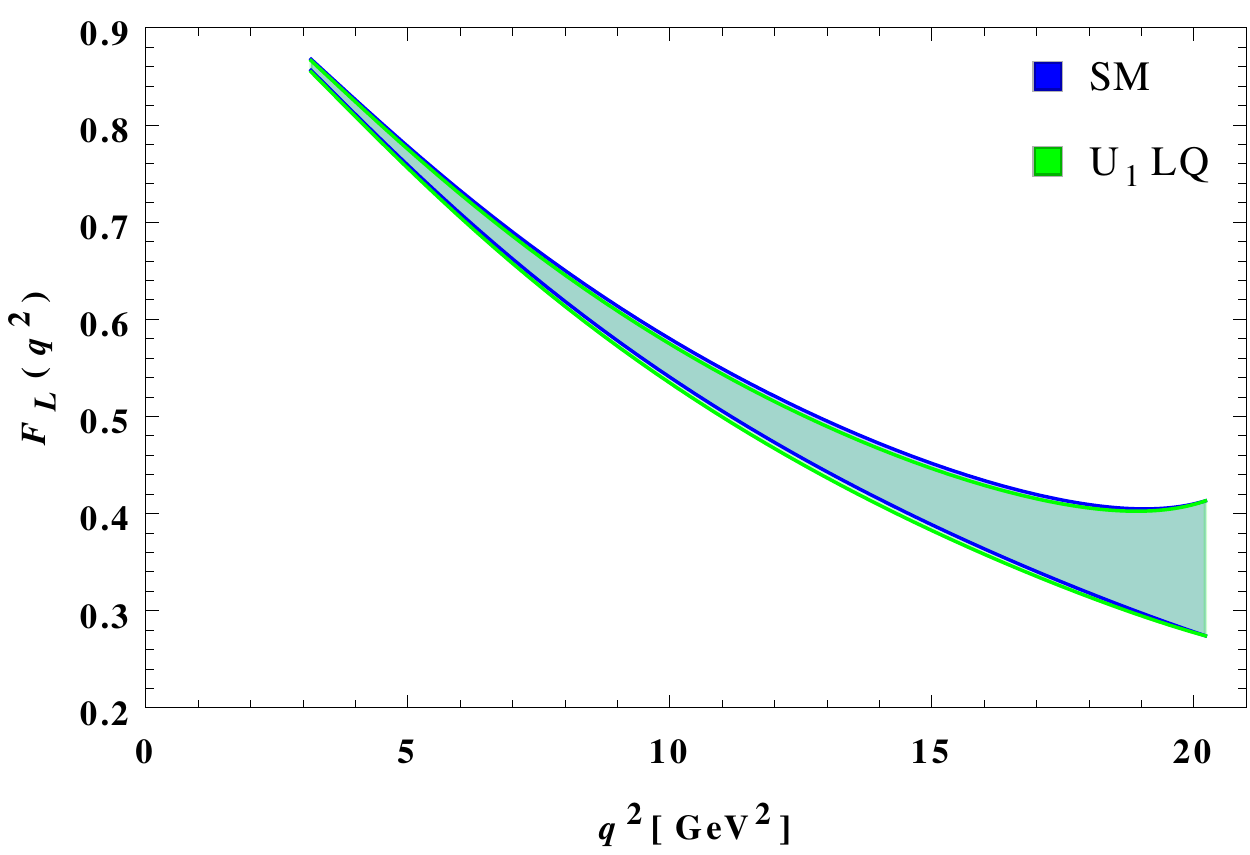} 
\caption{Variation of differential branching fraction, $R_\omega^{\tau/\ell}$, $P_\tau$, $A_{FB}$ and $F_L$  observables with  respect to $q^2$ for $B\to \omega\tau\bar{\nu}$ process.}
\label{B2omfig}
\end{figure}

The form-factors for $B\to (\rho, \omega)\tau\bar{\nu}$ decay are determined by light cone sum rule (LCSR) technique \cite{Straub:2015ica}, which are parametrized as
\begin{equation}
F_i(q^2) = \left(1-q^2/ m^2_{R,i}\right)^{-1} \sum_{k=0} a^i_{k} \left[z(q^2) -z(0)\right]^k, \label{form}
\end{equation} 
where $z(q^2) = \frac{\sqrt{t_+-q^2}-\sqrt{t_+-t_0}}{\sqrt{t_+-q^2}+\sqrt{t_+-t_0}}$, $t_{\pm} = (M_B\pm M_{\rho, \omega})^2$ and $t_0 = t_+(1-\sqrt{1-t_-/t_+})$. Here the form-factors $F_i$ refer to $V(q^2)$, $A_0(q^2)$, $A_1(q^2)$ and $A_{12}(q^2)$, where $A_{12}(q^2)$ is defined as
\begin{equation}
A_{12}(q^2) = \frac{\left(M_B + M_{\rho, \omega}\right)^2 \left(M^2_B -M^2_{\rho, \omega} - q^2\right) A_1(q^2) - \lambda_{\rho, \omega} A_2(q^2)}{16 M_B M^2_{\rho, \omega} (M_B+ M_{\rho, \omega})}.
\label{A12}
\end{equation}
The values of the resonance masses in Eq. (\ref{form}) are considered as  $m_{R,V} = 5.325$ GeV, $m_{R, A_0} = 5.279$ GeV, $m_{R, A_1} = 5.724$ GeV and $m_{R, A_{12}} = 5.724$ GeV for both the decays. The inputs of the form-factors for $B\to \rho$ decay are~\cite{Straub:2015ica}
\begin{eqnarray}
& & a^V_0 = 0.33(3), \quad a^V_1 = -0.86(18), \quad a^V_2 = 1.80(97), \quad a^{A_0}_0 = 0.36(4), \quad a^{A_0}_1 = -0.83(20),\nonumber\\
& & a^{A_0}_2 = 1.33(1.05), \quad a^{A_1}_0 = 0.26(3), \quad a^{A_1}_1 = 0.39(14), \quad a^{A_1}_2 = 0.16(41),\nonumber\\
& & a^{A_{12}}_0 = 0.30(3), \quad a^{A_{12}}_1 = 0.76(20), \quad a^{A_{12}}_2 = 0.46(76),
\end{eqnarray}
whereas those for $B\to \omega$ decay are~\cite{Straub:2015ica}
\begin{eqnarray}
& & a^V_0 = 0.30(4), \quad a^V_1 = -0.83(29), \quad a^V_2 = 1.72(1.24), \quad a^{A_0}_0 = 0.33(5), \quad a^{A_0}_1 = -0.83(30),\nonumber\\
& & a^{A_0}_2 = 1.42(1.25), \quad a^{A_1}_0 = 0.24(3), \quad a^{A_1}_1 = 0.34(24), \quad a^{A_1}_2 = 0.09(57),\nonumber\\
& & a^{A_{12}}_0 = 0.27(4), \quad a^{A_{12}}_1 = 0.66(26), \quad a^{A_{12}}_2 = 0.28(98).
\end{eqnarray}
We use these form-factors in our computation and calculate the branching fraction, $R^{\tau/\ell}_{\rho,\omega}$, $P_{\tau}$, $A_{FB}$ and $F_L$  observables for both the decays. In Figs.~\ref{B2rhofig} and \ref{B2omfig}, we plot these observables as a function of $q^2$ for $B\to \rho\tau\bar{\nu}$ and $B\to \omega\tau\bar{\nu}$ respectively. The computed  average values of these observables for both decays are  listed in Tables~\ref{B2rhotab} and \ref{B2omtab} respectively. Analogous to $B \to \pi \tau \bar \nu $ mode,  in this case also, i.e., for both the decay modes, the branching fractions as well as the lepton non-universality parameters $R_{\rho,\omega}^{\tau/\ell}$ deviate significantly from their SM predictions due to the LQ effect whereas the observables $P_\tau$, $A_{FB}$ and  $F_L$ are almost consistent with their SM estimations. 
\begin{table}[h!]
\begin{tabular}{|c|c|c|c|c|c|}
\hline\hline
 & ${\rm Br}(B\to \rho\tau\bar{\nu})$ & $R_{\rho}^{\tau/\ell}$ & $P_{\tau}$ & $A_{FB}$& $F_L$\\
 \hline
 SM &~  $(2.165\pm 0.545)\times 10^{-4}$ ~& ~$0.526\pm 0.021$ ~&~ $-0.540\pm 0.056$ ~&~ $-0.174\pm 0.060$ ~&~ $0.504\pm 0.086$~\\
 \hline
 $U_1$ LQ & $(3.277\pm 0.828)\times 10^{-4}$ & $0.796\pm 0.031$ & $-0.556\pm 0.053$& $-0.181\pm 0.060$ & $0.500\pm 0.087$\\
 \hline\hline
\end{tabular}
\caption{Predicted values of various observables for $B\to \rho\tau\bar{\nu}$ decay mode both in the SM as well as in LQ model. }
\label{B2rhotab}
\end{table}
\begin{table}[h!]
\begin{tabular}{|c|c|c|c|c|c|}
\hline\hline
 & ${\rm Br}(B\to \omega\tau\bar{\nu})$ & $R_{\omega}^{\tau/\ell}$ & $P_{\tau}$ & $A_{FB}$& $F_L$\\
 \hline
 SM &~  $(1.828\pm 0.554)\times 10^{-4}$ ~& ~$0.529\pm 0.031$ ~&~ $-0.535\pm 0.080$ ~&~ $-0.175\pm 0.085$ ~&~ $0.500\pm 0.120$~\\
 \hline
 $U_1$ LQ & $(2.765\pm 0.843)\times 10^{-4}$ & $0.800\pm 0.046$ & $-0.552\pm 0.075$& $-0.183\pm 0.085$ & $0.495\pm 0.122$\\
 \hline\hline
\end{tabular}
\caption{Predictions for various observables of $B\to \omega\tau\bar{\nu}$ decay in the SM as well as in the $U_1$ LQ model.}
\label{B2omtab}
\end{table}

\subsection{$B_s\to (K, K^*)\tau\bar{\nu}$ decays:}
\begin{figure}[h!]
\begin{tabular}{cc}
\includegraphics[width=70mm]{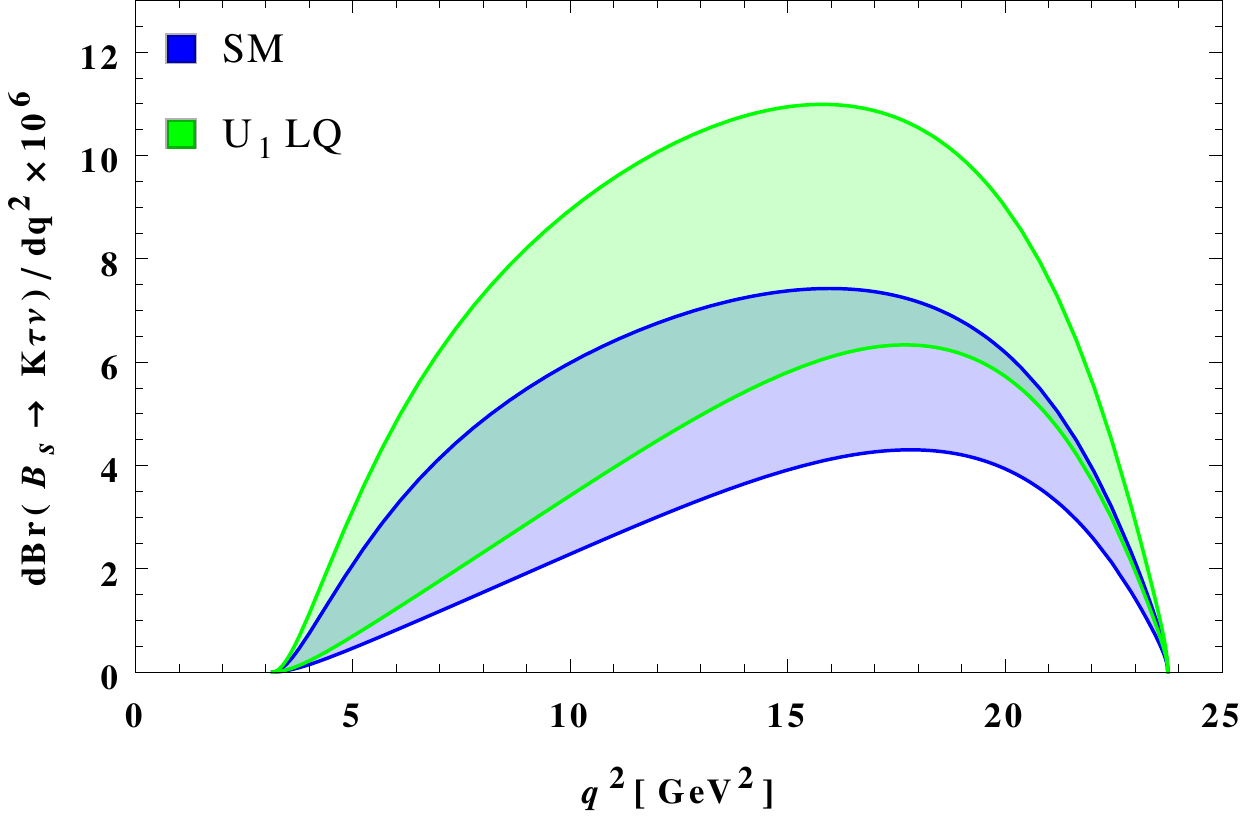} & \includegraphics[width=70mm]{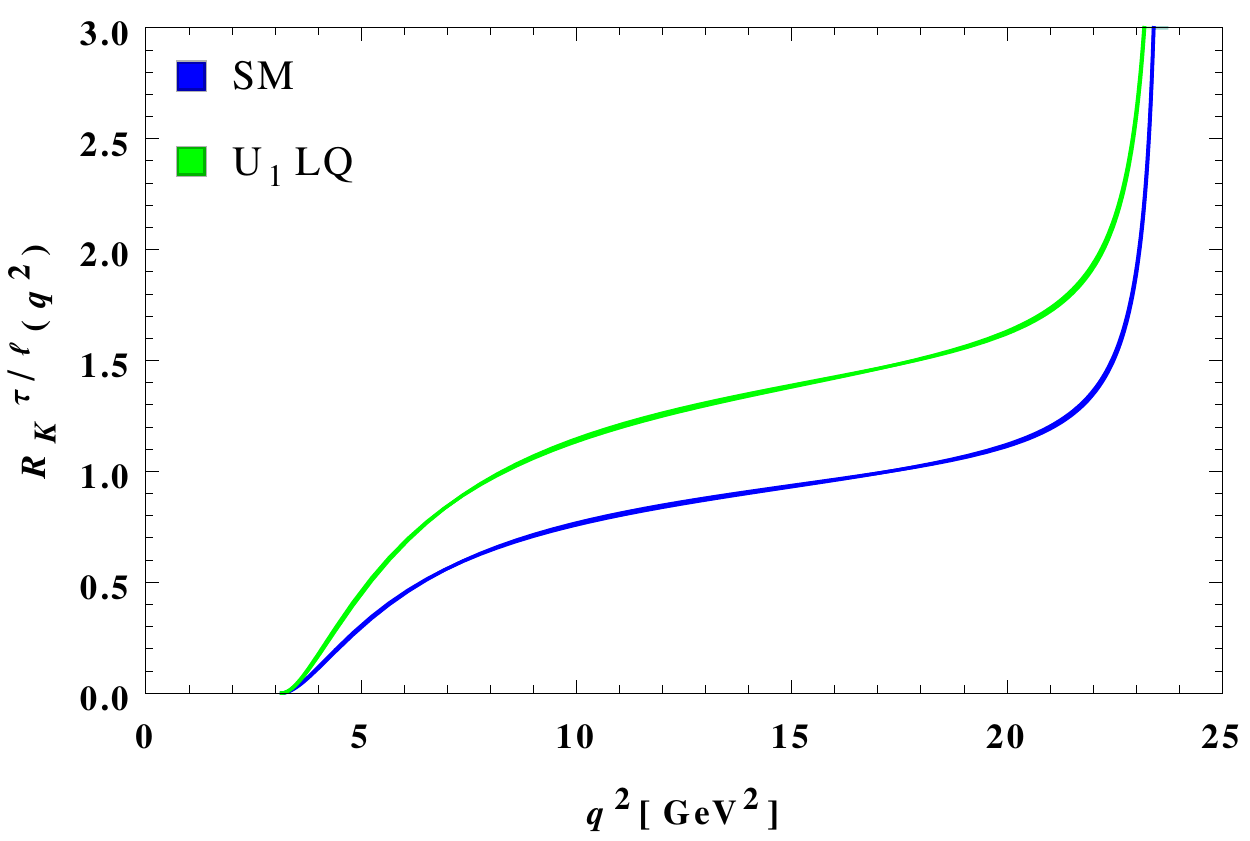} \\
\includegraphics[width=70mm]{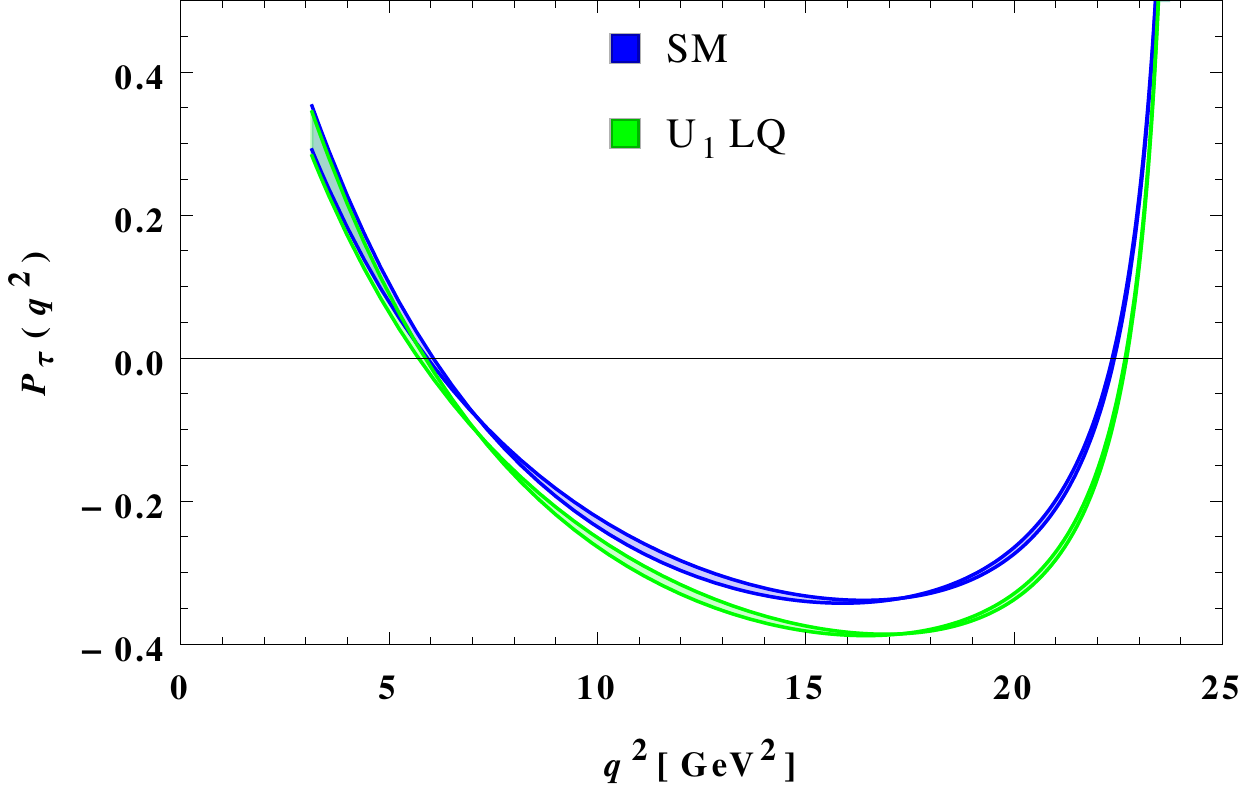} & \includegraphics[width=70mm]{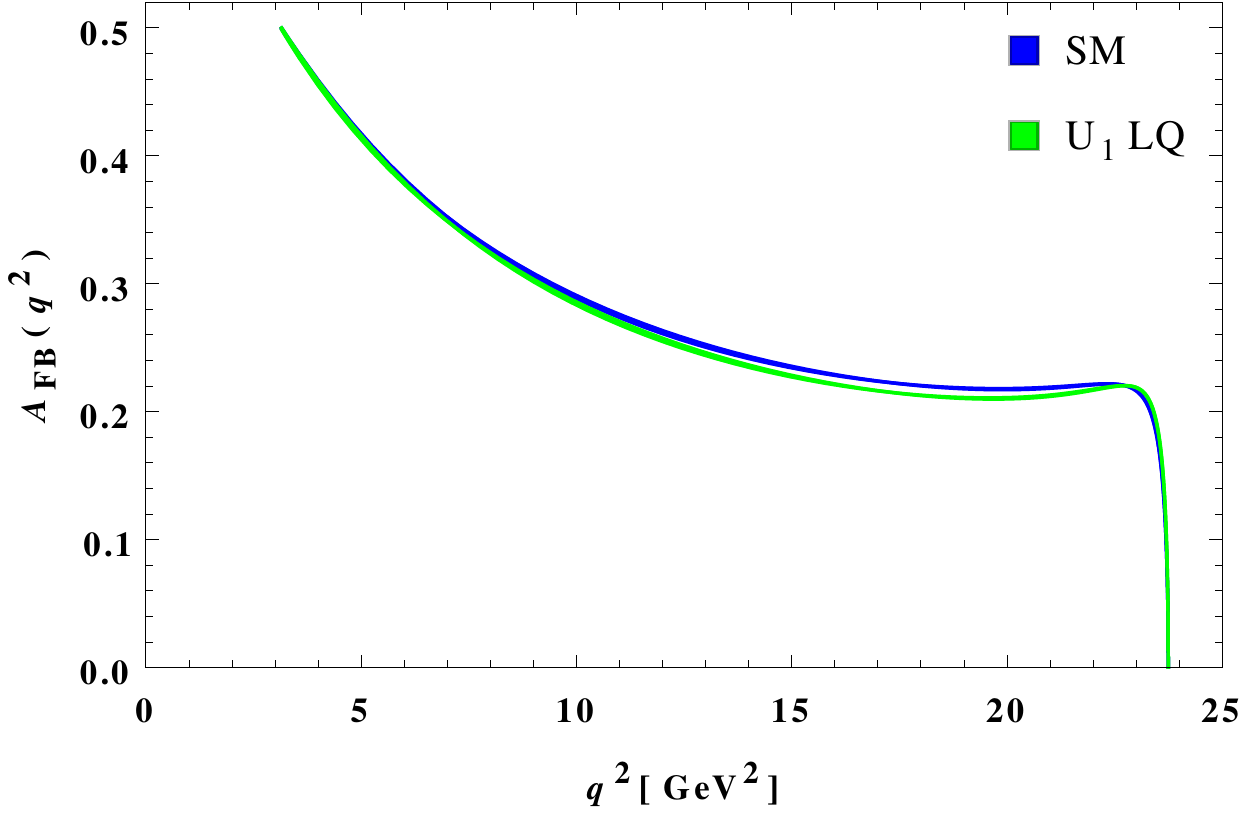} 
\end{tabular}
\caption{The $q^2$ variation plots for the branching fraction, $R^{\tau/\ell}_K$, $P_{\tau}$ and $A_{FB}$ of $B_s\to K\tau\bar{\nu}$ decay.}
\label{B2Kfig}
\end{figure}

The form-factors of $B_s \to K$ transition are determined in lattice QCD technique. In this approach, the two relevant form-factors are parametrized as follows~\cite{Bazavov:2019aom}
\begin{eqnarray}
F_+(q^2) &=& \left(1-q^2/m^2_{B^{*+}}\right)^{-1} \sum^{K-1}_{k=0} b^+_k \left[z^k - (-1)^{k-K} \frac{k}{K}~z^K\right], \nonumber \\
F_0(q^2) &=& \left(1-q^2/m^2_{B^{*0}}\right)^{-1} \sum^{K-1}_{k=0} b^0_k~ z^k,
\end{eqnarray}
where $z(q^2) = \frac{\sqrt{t_{\rm cut}-q^2}-\sqrt{t_{\rm cut}-t_0}}{\sqrt{t_{\rm cut}-q^2}+\sqrt{t_{\rm cut}-t_0}}$, $m_{B^{*+}} = 5.32465$ GeV, $m_{B^{*0}} = 5.68$ GeV, $\sqrt{t_{\rm cut}} = 5.414$ GeV, $t_0 = t_{\rm cut}-\sqrt{t_{\rm cut}\left(t_{\rm cut} -t_-\right)}$ and $t_-= (M_{B_s}-M_K)^2$.
The values of the input parameters in the above mentioned form-factors are as follows~\cite{Bazavov:2019aom}
\begin{eqnarray}
& & b^+_0 = 0.3623(0.0178),\quad b^+_1 = -0.9559(0.1307), \quad b^+_2 = -0.8525(0.4783), \nonumber \\
& & b^+_3 = 0.2785(0.6892), \quad b^0_0 = 0.1981(0.0101), \quad b^0_1 = -0.1661(0.1130),\nonumber \\
& & b^0_2 = -0.6430(0.4385), \quad b^0_3 = -0.3754(0.4535).
\end{eqnarray}
With these values, we calculate the branching fraction, $R^{\tau/\ell}_{K}$, $P_{\tau}$ and $A_{FB}$ of $B_s\to K\tau\bar{\nu}$ decay for the SM and for the $U_1$ LQ model. In Fig.~\ref{B2Kfig}, we plot these quantities as a function of $q^2$ and also listed their  predicted values  in Table~\ref{B2Ktab}. In this case the branching fraction and the $P_\tau$ observables have mild deviation from their SM values due to the effect of $U_1$ LQ whereas discrepancy between between SM and LQ predicted values for $R_K^{\tau/l}$ observable is considerably large. On the other hand the forward-backward asymmetry parameter remains consistent with its SM value in the LQ scenario.
\begin{table}[h!]
\begin{tabular}{|c|c|c|c|c|}
\hline\hline
 & ${\rm Br}(B_s\to K\tau\bar{\nu})$ & $R_{K}^{\tau/\ell}$ & $P_{\tau}$ & $A_{FB}$\\
 \hline
 SM &  $~(0.765\pm 0.155)\times 10^{-4}$~ &~ $0.767\pm 0.073$ ~&~ $-0.244\pm 0.060$ ~&~ $0.253\pm 0.007$~\\
 \hline
 $U_1$ LQ & $(1.129\pm 0.230)\times 10^{-4}$ & $1.133\pm 0.104$ & $-0.290\pm 0.057$& $0.248\pm 0.008$\\
 \hline\hline
\end{tabular}
\caption{Predicted values of the observables for $B_s\to K\tau\bar{\nu}$ decay process in both the SM and the LQ model.}
\label{B2Ktab}
\end{table}

\begin{figure}[h!]
\begin{tabular}{cc}
\includegraphics[width=70mm]{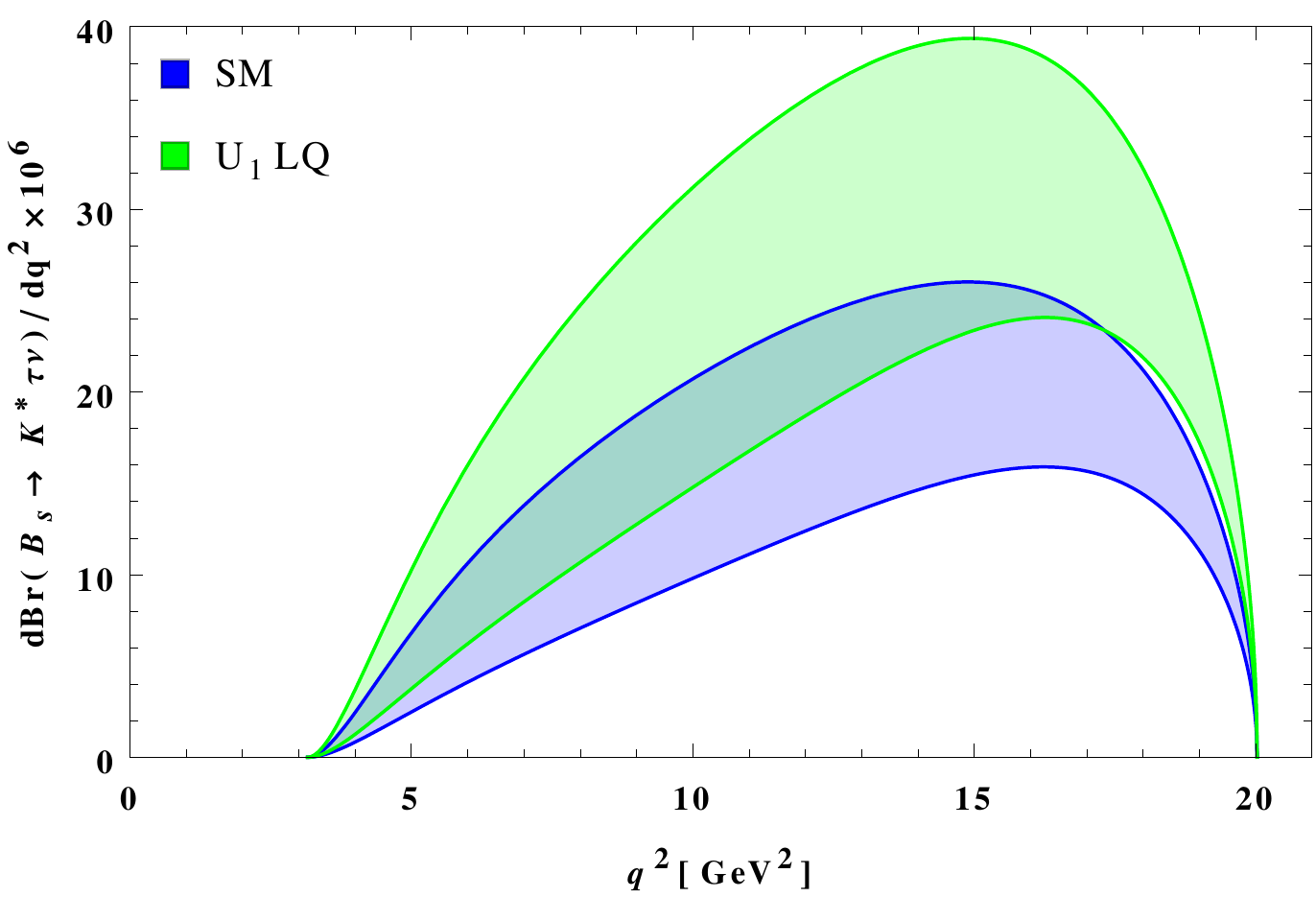} & \includegraphics[width=70mm]{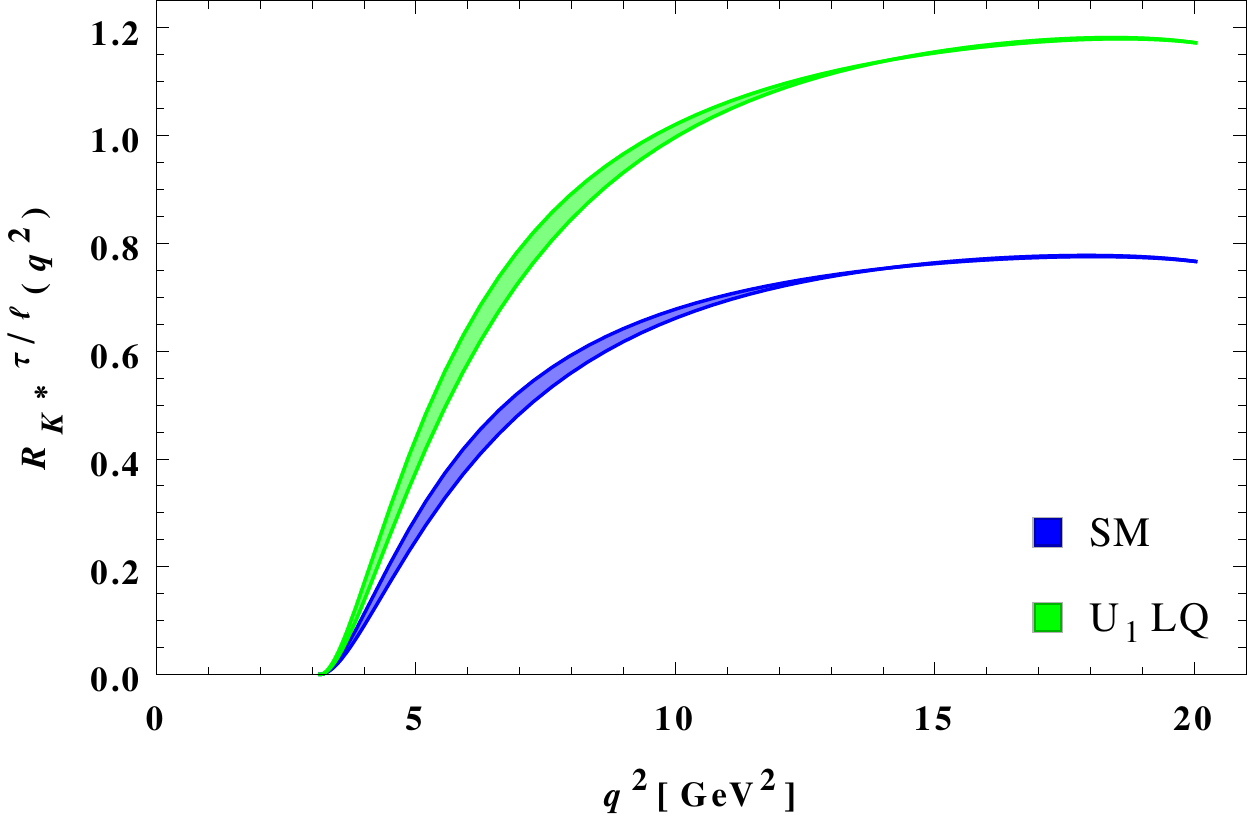} \\
\includegraphics[width=70mm]{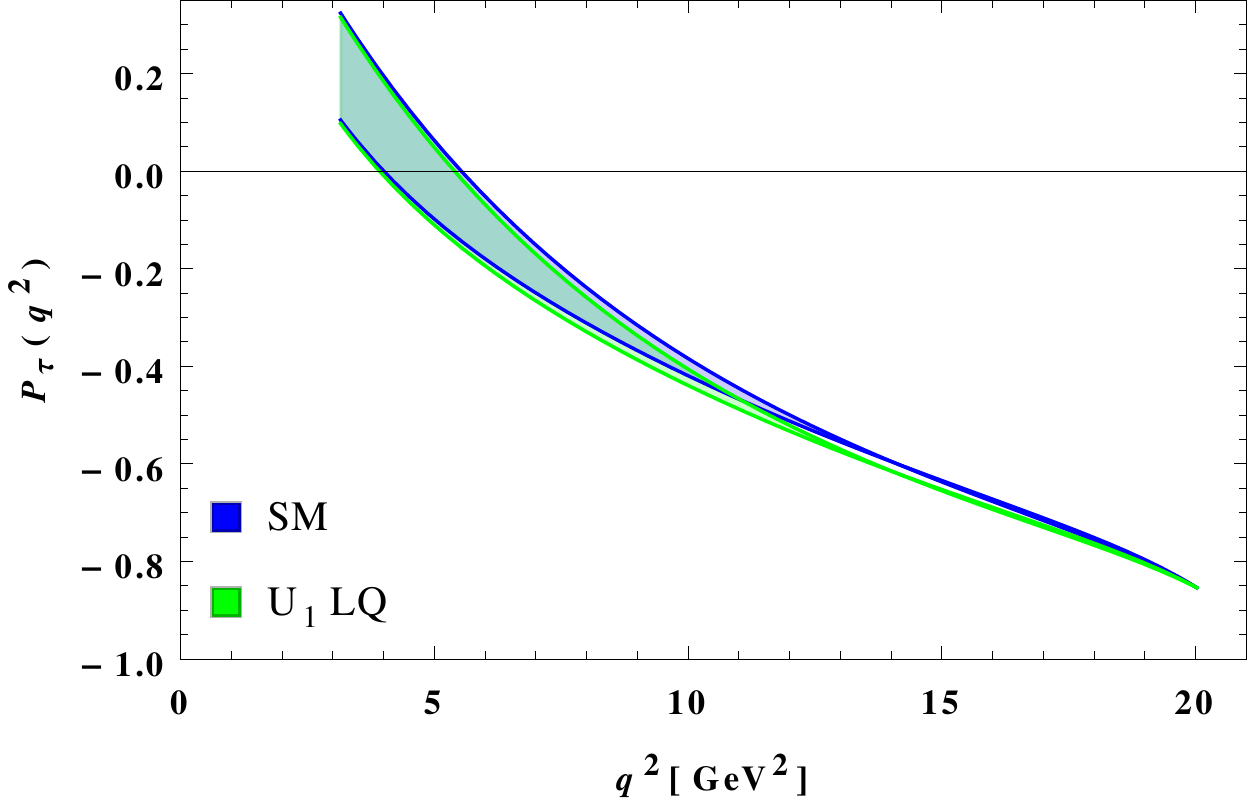} & \includegraphics[width=70mm]{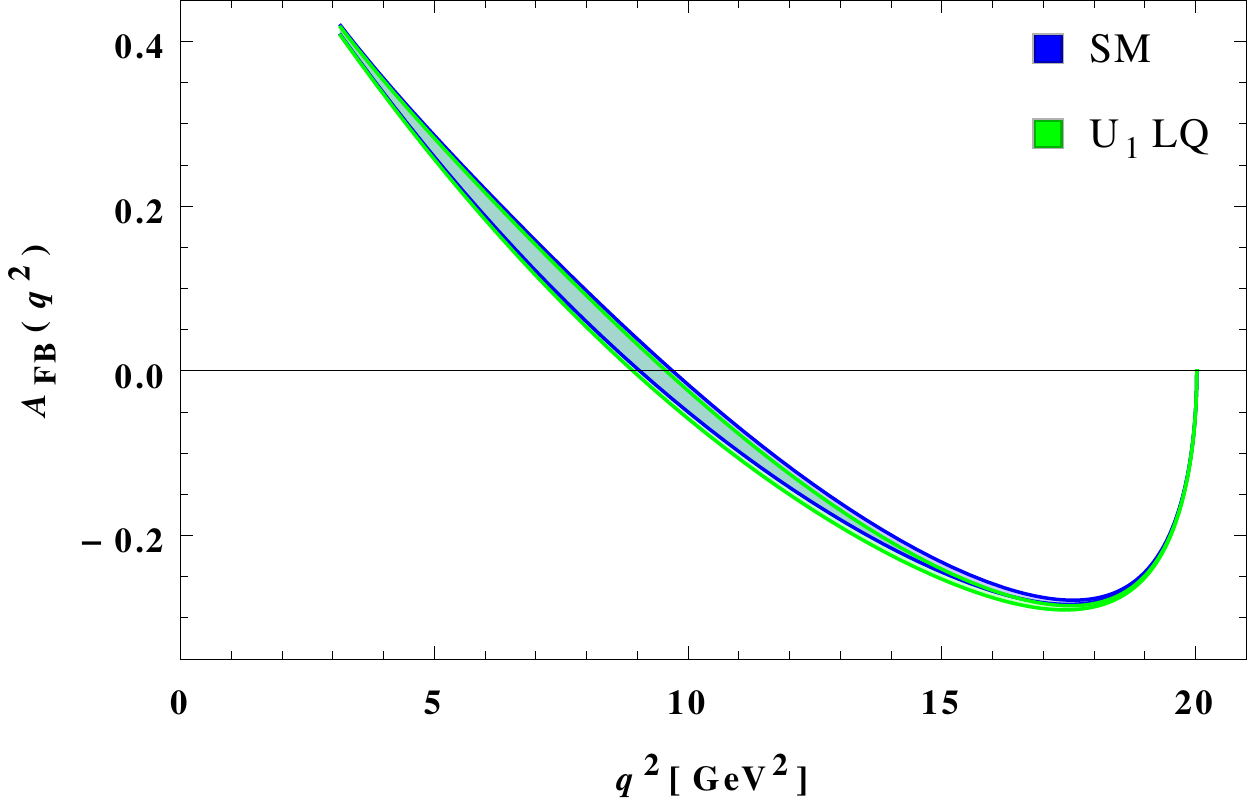} 
\end{tabular}
\includegraphics[width=70mm]{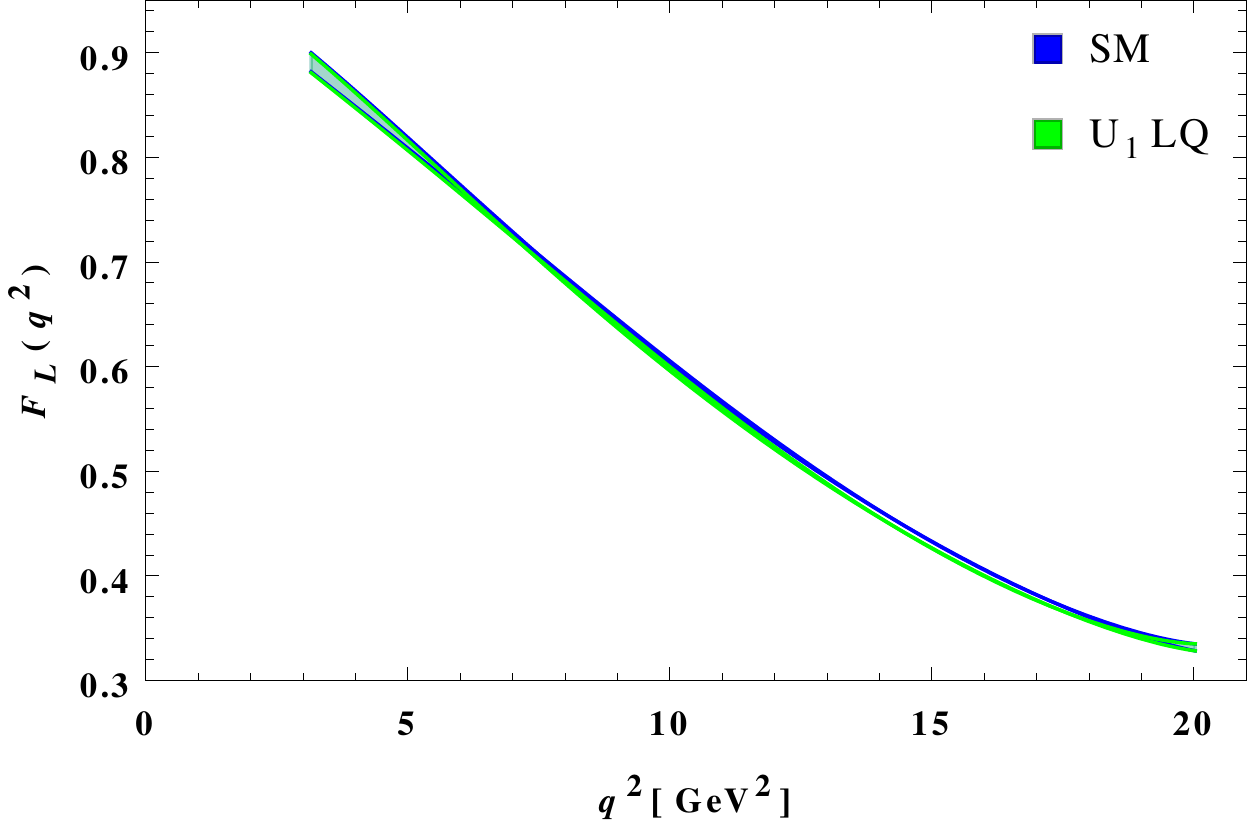} 
\caption{The $q^2$ variation plots for the branching fraction, $R^{\tau/\ell}_{K^*}$, $P_{\tau}$, $A_{FB}$ and $F_L$ of $B_s\to K^*\tau\bar{\nu}$ decay process.}
\label{B2Ksfig}
\end{figure}

For $B_s\to K^*\tau\bar{\nu}$ decay process, we use the form-factors    calculated using lattice QCD approach, which are expressed as~\cite{Horgan:2013hoa}
\begin{equation}
F(q^2) = \frac{1}{P(q^2;\Delta m)}\left[a_0 +a_1 z(q^2)\right],
\end{equation}
where $F$ refers to the form-factors $V(q^2)$, $A_0(q^2)$, $A_1(q^2)$ and $A_{12}(q^2)$. The expression of $A_{12}(q^2)$ is the same as that of Eq.~\ref{A12}. Here $P(q^2; \Delta m) = 1- q^2/\left(M_{B_s}+\Delta m\right)^2$ where $\Delta m =-87$ MeV for $A_0(q^2)$, $\Delta m =-42$ MeV for $V(q^2)$ and $\Delta m =350$ MeV for $A_1(q^2)$ and $A_{12}(q^2)$. The expansion parameter is defined as $z(q^2) = \frac{\sqrt{t_+ -q^2}-\sqrt{t_+ - t_0}}{\sqrt{t_+ -q^2}+\sqrt{t_+ - t_0}}$, where $t_0 = 12$ GeV and $t_{\pm} = \left(M_{B_s}\pm M_{K^*}\right)^2$. The input parameters of the form-factors are given by~\cite{Horgan:2013hoa}
\begin{eqnarray}
&& a_0^V = 0.322(0.048), \quad a_1^V = -3.04(0.67), \quad a_0^{A_0} = 0.476 (0.042), \nonumber \\ 
&& a_1^{A_0} = -2.29(0.74), \quad a_0^{A_1} = 0.2342(0.0122), \quad a_1^{A_1} = 0.100(0.174), \nonumber \\
& & a_0^{A_{12}} = 0.1954 (0.0133), \quad a_1^{A_{12}} = 0.350(0.190).
\end{eqnarray}
We calculate the branching fraction, $R^{\tau/\ell}_{K^*}$, $P_{\tau}$, $A_{FB}$ and $F_L$ for $B_s\to K^*\tau\bar{\nu}$ decay in the SM as well as in the $U_1$ LQ model. We plot these observables as a function of $q^2$ as shown in Fig.~\ref{B2Ksfig}. We also compute their average values  and list them in Table~\ref{B2Kstab}. For this process also the LQ effect is significant only for branching fraction and the lepton non-universality parameter $R_{K^*}^{\tau/\ell}$.
\begin{table}[h!]
\begin{tabular}{|c|c|c|c|c|c|}
\hline\hline
 & ${\rm Br}(B_s\to K^*\tau\bar{\nu})$ & $R_{K^*}^{\tau/\ell}$ & $P_{\tau}$ & $A_{FB}$& $F_L$\\
 \hline
 SM &~  $(2.259\pm 0.449)\times 10^{-4}$~ &~ $0.580\pm 0.023$ ~&~ $-0.534\pm 0.043$ ~&~ $-0.135\pm 0.040$ ~&~ $0.505\pm 0.039$~\\
 \hline
 $U_1$ LQ & $(3.416\pm 0.680)\times 10^{-4}$ & $0.877\pm 0.033$ & $-0.552\pm 0.040$& $-0.142\pm 0.040$ & $0.500\pm 0.039$\\
 \hline\hline
\end{tabular}
\caption{Predictions for the observables in $B\to K^*\tau\bar{\nu}$ decay process in the SM as well as in the LQ model.}
\label{B2Kstab}
\end{table}

\section{Conclusions}
Probing the extension of the SM  at the TeV scale is one of the prime goals of LHC experiment. However, in the absence of any direct observation of NP signal at LHC, we need to adopt alternative strategies. In this context, the results LHCb and $B$ factory experiments may be examined seriously to look for any smoking-gun signal of NP beyond the SM. In fact, the recent observation of various flavour anomalies associated with $b \to c \ell \bar \nu$ and $b \to s \ell^+ \ell^-$ transitions  may be considered as one of the most  imperative hints of NP at the TeV scale. However, it is really a challenging task to explain these appealing set of anomalies in a coherent manner using a single platform, as the NP scales involved in the CC and NC sectors differ significantly. There are only a handful of models which can provide simultaneous solutions  to the discrepancies of both these sectors. The vector LQ model, where the SM is extended by an additional  TeV scale LQ $U_1(3,1,2/3)$ is known  to be one such model. Therefore, in this work we have performed a detailed study of the impact of the  $U_1$ LQ on the rare semileptonic decay channels mediated by $b \to u \tau \bar \nu$ transitions.
To constrain the new physics parameters we have performed a global fit using various observables in the $b \to c \ell \bar \nu$, $b \to s \ell^+ \ell^-$ as well as the  $b \to u \tau \bar \nu$ transitions which show few sigma deviations. 
After ensuring that we are dealing  with scenarios allowed by $b \to c \ell \bar{\nu}$  as well as $b \to s \ell^+ \ell^- $ anomalies,  we made the predictions for different observables
of $B \to (\pi, \rho, \omega) \tau \bar \nu$ as well as $B_s \to (K, K^*)\tau \bar \nu$ processes. The list of these observables include   branching fractions,  lepton non-universality parameters, forward backward asymmetries, lepton polarization asymmetries as well as longitudinal polarization of the final vector mesons.  We found that in all these processes the branching fractions  as well as the lepton non-universality parameters $R_{P,V}^{\tau/\ell}$ show significant deviation from their corresponding SM predictions whereas the impact of $U_1$ LQ on other observables is rather mild. Since, the observables $R_{P,V}^{\tau/\ell}$ are fairly clean, i.e., essentially free from hadronic uncertainties, with large deviations from their SM values, it is strongly urged to search for them in the LHCb or Belle II experiments. If such observables are measured, they would provide an indirect signal for the possible existence of TeV scale vector LQ.

\acknowledgments

We thank the organizers of WHEPP 2019 at IIT Guwahati, where this work was  initiated. RM  acknowledges the support from  SERB, Government of India, through grant No. EMR/2017/001448.

\end{document}